 \documentclass[]{jfm_arxiv}

\usepackage{graphicx}
\usepackage{newtxtext}
\usepackage{newtxmath}
\usepackage{natbib}
\usepackage{hyperref}
\usepackage{comment}
\hypersetup{
    colorlinks = true,
    urlcolor   = blue,
    citecolor  = black,
}
\usepackage[usenames, dvipsnames]{xcolor}

\DeclareFontEncoding{LS1}{}{}
\DeclareFontSubstitution{LS1}{stix}{m}{n}
\DeclareFontEncoding{LS2}{}{\noaccents@}
\DeclareFontSubstitution{LS2}{stix}{m}{n}
\DeclareSymbolFont{SYMB}{LS1}{stixscr}{m}{n}
\DeclareSymbolFont{SYMB4}{LS1}{stixbb}{m}{it}
\DeclareSymbolFont{ARR3}{LS2}{stixtt}{m}{n}
\DeclareMathSymbol{\pentagonblack}{\mathord}{SYMB4}{"DA}
\DeclareMathSymbol{\bigblacktriangleup}{\mathord}{SYMB}{"C9}
\DeclareMathSymbol{\bigblacktriangledown}{\mathord}{SYMB}{"D3}
\DeclareMathSymbol{\mdlgblkdiamond}{\mathord}{SYMB}{"DD}
\DeclareMathSymbol{\mdlgwhtdiamond}{\mathord}{SYMB}{"DE}
\DeclareMathSymbol{\circletophalfblack}{\mathord}{SYMB}{"EA}
\DeclareMathSymbol{\circlerighthalfblack}{\mathord}{SYMB}{"E8}
\DeclareMathSymbol{\circlebottomhalfblack}{\mathord}{SYMB}{"E9}
\DeclareMathSymbol{\squaretopblack}{\mathord}{SYMB4}{"CD}
\DeclareMathSymbol{\squarebotblack}{\mathord}{SYMB4}{"CE}
\DeclareMathSymbol{\mdwhtcircle}{\mathord}{ARR3}{"7D}
\DeclareMathSymbol{\mdwhtsquare}{\mathord}{ARR3}{"9A}

\definecolor{darkviolet}{rgb}{0.58, 0.0, 0.83}
\definecolor{shamrockgreen}{rgb}{0.0, 0.62, 0.38}
\definecolor{deepskyblue}{rgb}{0.0, 0.75, 1.0}
\definecolor{amber}{rgb}{1.0, 0.49, 0.0}
\definecolor{red}{rgb}{1.0, 0.0, 0.0}

\newcommand\solid[1][1.0cm]{\rule[0.4ex]{#1}{2.8pt}}

\newcommand\dashdot{\mbox{%
  \solid[3mm]\hspace{1.mm}\solid[1mm]\hspace{1.mm}\solid[3mm]}}

\newcommand\dashedl{\mbox{%
  \solid[1.5mm]\hspace{1.mm}\solid[1.5mm]\hspace{1.mm}\solid[1.5mm]\hspace{1.mm}\solid[1.5mm]}}
\newcommand\dottedl{\mbox{%
  \solid[1mm]\hspace{1.mm}\solid[1mm]\hspace{1.mm}\solid[1mm]\hspace{1.mm}\solid[1mm]\hspace{1.mm}\solid[1mm]
}}

\newcommand{\diff}{\mathrm{d}}

 \title[Streamwise velocity variance asymptotics]{On the streamwise velocity variance in the near-wall region of turbulent flows}

\author{Sergio Pirozzoli\aff{1} 
 \corresp{\email{sergio.pirozzoli@uniroma1.it}}}

\affiliation{
\aff{1} Dipartimento di Ingegneria Meccanica e Aerospaziale, Sapienza Universit\`a di Roma, Via Eudossiana 18, 00184 Roma, Italy
}

\date{\today}

\begin{document}

\maketitle

\begin{abstract}
We study the behaviour of the streamwise velocity variance in turbulent wall-bounded flows using a DNS
database of pipe flow up to $\Rey_{\tau} \approx 12000$. The analysis of the spanwise spectra
in the viscous near-wall region strongly hints to the presence of an overlap layer
between the inner- and the outer-scaled spectral ranges, featuring a $k_{\theta}^{-1+\alpha}$ decay (with $k_{\theta}$ the wavenumber in the azimuthal direction, and $\alpha \approx 0.18$), 
hence shallower than suggested by the classical formulation of the attached-eddy model. 
The key implication is that the contribution to the streamwise velocity variance from the largest scales of motion
(superstructures) slowly declines as $\Rey_{\tau}^{-\alpha}$, and the integrated
variance follows a defect power law of the type $\left< u^2 \right>^+ = A - B \, \Rey_{\tau}^{-\alpha}$,
with constants $A$ and $B$ depending on $y^+$.
The DNS data very well support this behaviour, which implies that strict 
wall scaling is restored in the infinite Reynolds number limit. 
The extrapolated limit distribution of the streamwise velocity variance features 
a buffer-layer peak value of $\left< u^2 \right>^+ \approx 12.1$, and an additional
outer peak with larger magnitude.
The analysis of the velocity spectra also suggests a similar behaviour of the dissipation rate of the streamwise velocity
variance at the wall, which is expected to attain a limiting value of about $0.28$, 
hence slightly exceeding the value $0.25$ which was assumed in previous analyses~\citep{chen_21}.
We have found evidence suggesting that the reduced near-wall influence of wall-attached eddies 
is likely linked to the formation of underlying turbulent Stokes layers.
\end{abstract}

\section{Introduction}

A fundamental challenge in fluid dynamics research is the identification of appropriate scaling laws for the turbulence properties.
In wall turbulence, the classical scaling for the velocity fluctuations is based on the friction
velocity, namely $u_{\tau} = (\tau_w/\rho)^{1/2}$, where $\tau_w$ is the wall shear stress, and $\rho$ is the fluid
density. Two length scales can instead be identified, namely the viscous (inner) length scale $\delta_v = \nu / u_{\tau}$
(with $\nu$ the fluid kinematic viscosity), and the outer length scale, say $\delta$, connected with the 
global dimensions of the wall layer.
This classical scenario, first envisaged by \citet{prandtl_25} was challenged by later 
experimental and numerical data~\citep[][to mention a few]{spalart_88,degraaff_00,metzger_01,marusic_01}, 
which showed that the inner-scaled wall-parallel velocity variances have a clear Reynolds number dependence.
A theoretical layout to explain violations from strict inner-layer universality was
set by~\citet{townsend_76}, through the so-called attached-eddy model, hereafter referred to as AEM.
The model appeals to the existence of a self-similar hierarchy of eddies rooted at the wall, which on account of the
impermeability condition can only convey wall-parallel velocity fluctuations in the wall proximity,
thus retaining a footprint which manifests itself with superposition and modulation effects~\citep{hutchins_07a,mathis_09a}.
The AEM and its subsequent extensions~\citep{perry_82,perry_86,perry_95,marusic_19} 
currently constitute the most complete theoretical
framework to explain the distributions of the statistical properties in wall turbulence.
One key prediction of the AEM is that the inner-scaled wall-parallel velocity variances at a fixed 
outer-scaled location should decrease logarithmically with the wall distance~\citep{townsend_76,perry_82,meneveau_13}.
A weaker corollary of the model~\citep{marusic_17,baars_20a} is that the inner-scaled velocity variances at fixed $y^+$ (hereafter, the '+' superscript refers to inner normalization) in the inner part of the wall layer
should increase logarithmically with $\Rey_{\tau}$, where $\Rey_{\tau} = \delta/ \delta_v = u_{\tau} \delta / \nu$, is the
friction Reynolds number.
If true, this corollary would imply that strict wall scaling is violated.

At least one potential weakness may be envisaged in applying the AEM to asymptotically high $\Rey_{\tau}$.
If the growth of the buffer-layer peak of the streamwise velocity variance were to persist indefinitely, 
and if the peak consistently occurred at the same position \citep[$y^+ \approx 15$, see][]{sreenivasan_89}, 
it would lead to an increasingly high probability of instantaneous negative velocity events. 
This would likely alter the nature of the flow.
Whereas the possibility of instantaneous velocity reversal in the viscous sublayer is known~\citep{lenaers_12},
it is hard to believe that this can extend to the buffer layer.
In fact, an alternative scenario has been recently advocated by \citet{chen_21}, whereby growth of the 
buffer-layer peak would saturate on account of a bound on the dissipation rate of the
streamwise velocity variance. This would in turn imply that the wall-parallel velocity variances follow 
a defect power-law dependence with the wall distance, rather than logarithmic. 
Eventually, strict wall scaling would be restored in the limit of very high Reynolds numbers. 
Although the model of \citet{chen_21} lacks at the moment solid mathematical foundations,
it nevertheless seems to comply with existing DNS and experimental data at least as well as the classical AEM.
Hence, it has stirred the interest of the community, 
stimulating a number of follow-up studies~\citep{chen_22,klewicki_22,monkewitz_22,nagib_22,hwang_23,monkewitz_23}.
A common conclusion drawn from those studies appears to be that discerning alternative 
scaling based solely on inspection of basic statistics, such as velocity variances, 
requires access to Reynolds numbers that are well beyond the capabilities of current 
and possibly future experimental and numerical approaches.

The key objective of this paper is showing that some insight into the asymptotic behaviour of 
turbulence in the near-wall region can in fact be achieved even with current day data.
For this purpose, we examine the velocity spectra obtained from a newly generated DNS dataset of turbulent pipe flow, 
reaching Reynolds numbers up to $\Rey_{\tau} \approx 12000$, based on which we believe that informed extrapolation is possible. 

\section{The DNS database}

\begin{table}
 \centering
\begin{tabular*}{1.\textwidth}{@{\extracolsep{\fill}}lcccccccc}
 Flow case & $L_z/R$ & Mesh ($N_\theta \times N_r \times N_z$) & $\Rey_b$ & $f$ & $\Rey_{\tau}$ & $T/\tau_t$ & Line style \\
 \hline
 B   & 15  & $768   \times  96  \times   768$ & $17000$  & $0.02719$ &  $495.6$  & 192.9 & \color{shamrockgreen}\solid \\
 C   & 15  & $1792  \times 164  \times  1792$ & $44000$  & $0.02119$ &  $1132.2$ & 50.4 & \color{deepskyblue}\solid \\
 C-FY& 15  & $1792  \times 328  \times  1792$ & $44000$  & $0.02122$ &  $1132.2$ & 46.1 & \color{deepskyblue}\dashedl \\
 C-L & 30  & $1792  \times 164  \times  3584$ & $44000$  & $0.02119$ &  $1132.3$ & 52.8 & \color{deepskyblue}\dottedl \\
 C-LL& 45  & $1792  \times 164  \times  5376$ & $44000$  & $0.02114$ &  $1131.0$ & 45.3 & \color{deepskyblue}\dashdot \\
 D   & 15  & $3072  \times 243  \times  3072$ & $82500$  & $0.01828$ &  $1972.0$ & 45.1 & \color{amber}\solid \\
 E   & 15  & $4608  \times 327  \times  4608$ & $133000$ & $0.01657$ &  $3026.8$ & 26.9 & \color{red}\solid \\
 F   & 15  & $9216  \times 546  \times  9216$ & $285000$ & $0.01421$ &  $6006.4$ & 18.2 & \color{darkviolet}\solid \\
 G   & 15  & $18432 \times 1024 \times 18432$ & $612000$ & $0.01242$ & $12054.5$ & 6.99 & \solid \\
 \hline
\end{tabular*}
\caption{Flow parameters for DNS of pipe flow.
$R$ is the pipe radius, $L_z$ is the pipe axial length,
$N_{\theta}$, $N_r$ and $N_z$ are the number of grid points in the 
azimuthal, radial and axial directions, respectively, 
$\Rey_b = 2 R u_b / \nu$ is the bulk Reynolds number,
$f = 8 \tau_w / (\rho u_b^2)$ is the friction factor,
$\Rey_{\tau} = u_{\tau} R / \nu$ is the friction Reynolds number,
$T$ is the time interval used to collect the flow statistics,
and $\tau_t = R/u_{\tau}$ is the eddy turnover time.
}
\label{tab:runs}
\end{table}

This paper extends upon a previous publication on the subject, 
where Reynolds numbers up to $\Rey_{\tau} \approx 6000$ were achieved~\citep{pirozzoli_21}.
Here, the database is enlarged and improved, by 
extending the time interval of previous DNS, and including a new data point at $\Rey_{\tau} \approx 12000$.
A list of the flow cases is reported in table~\ref{tab:runs}, which includes basic information about the computational mesh
and some key parameters. The numerical algorithm is the same as in \citet{pirozzoli_21}, and details on the
mesh resolution are provided in appendix~\ref{sec:gridsens}.
As one can see in table~\ref{tab:runs}, the largest DNS has run for less than ten eddy turnover
times, which is the commonly accepted limit to guarantee time convergence~\citep{hoyas_06}.
Nevertheless, careful examination of the time convergence according to the method
of \citet{russo_17} has shown that the estimated standard deviation in the prediction of the streamwise velocity variance 
in the range of wall distances under scrutiny here ($y^+ \le 400$), is at most $0.6\%$.
Uncertainty is obviously larger in the velocity spectra, for which confidence bands are provided, see e.g. figure~\ref{fig:specz_inner}.
Essential details regarding the mean velocity profiles are provided in appendix~\ref{sec:statistics}; 
however, a comprehensive overview of the DNS results will be presented in future publications.
Here, the emphasis is on the spectra of streamwise velocity and the associated variances.

\section{Flow organization}

\begin{figure}
\centering
B~\includegraphics[width=4.0cm]{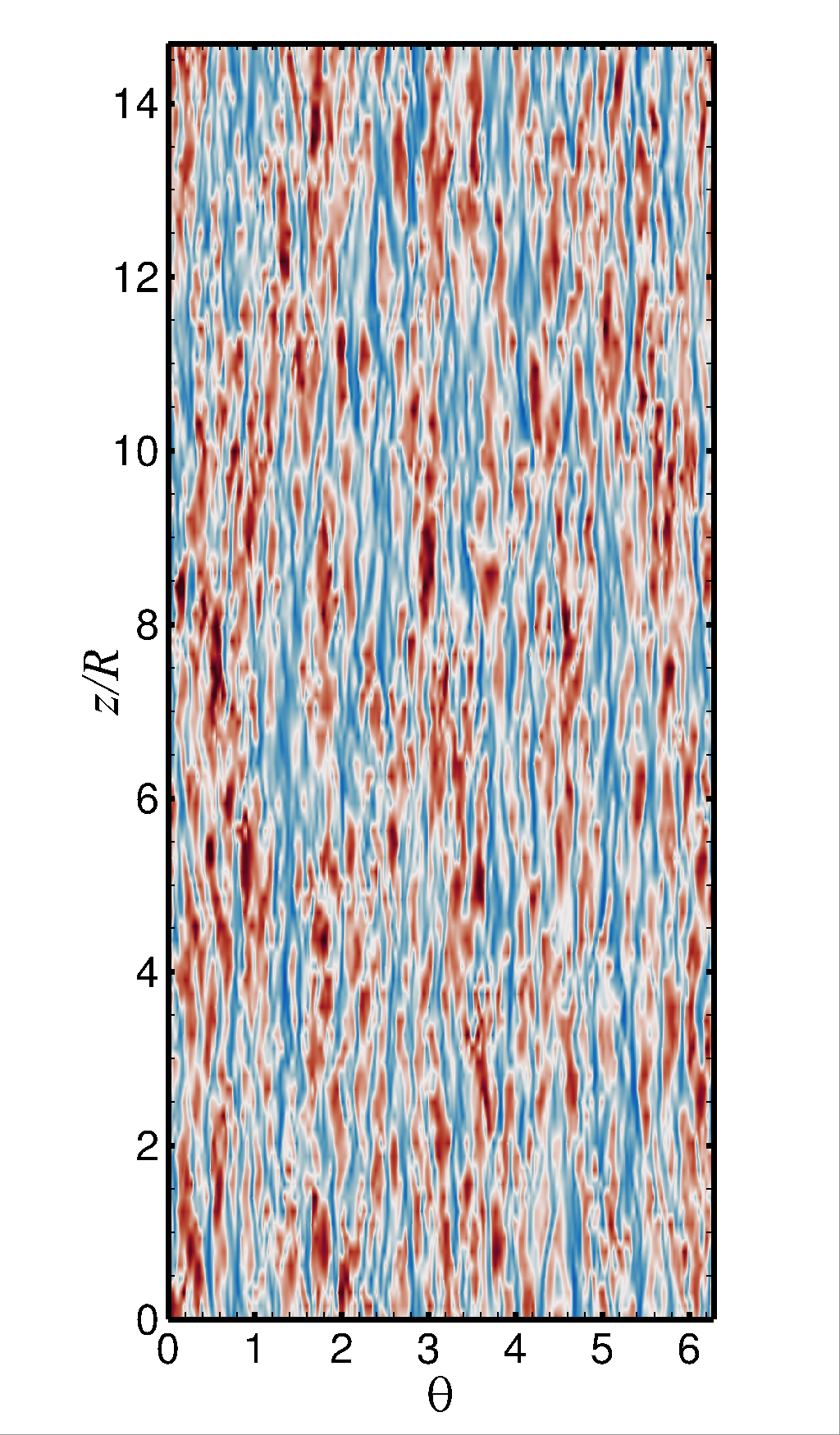}
C~\includegraphics[width=4.0cm]{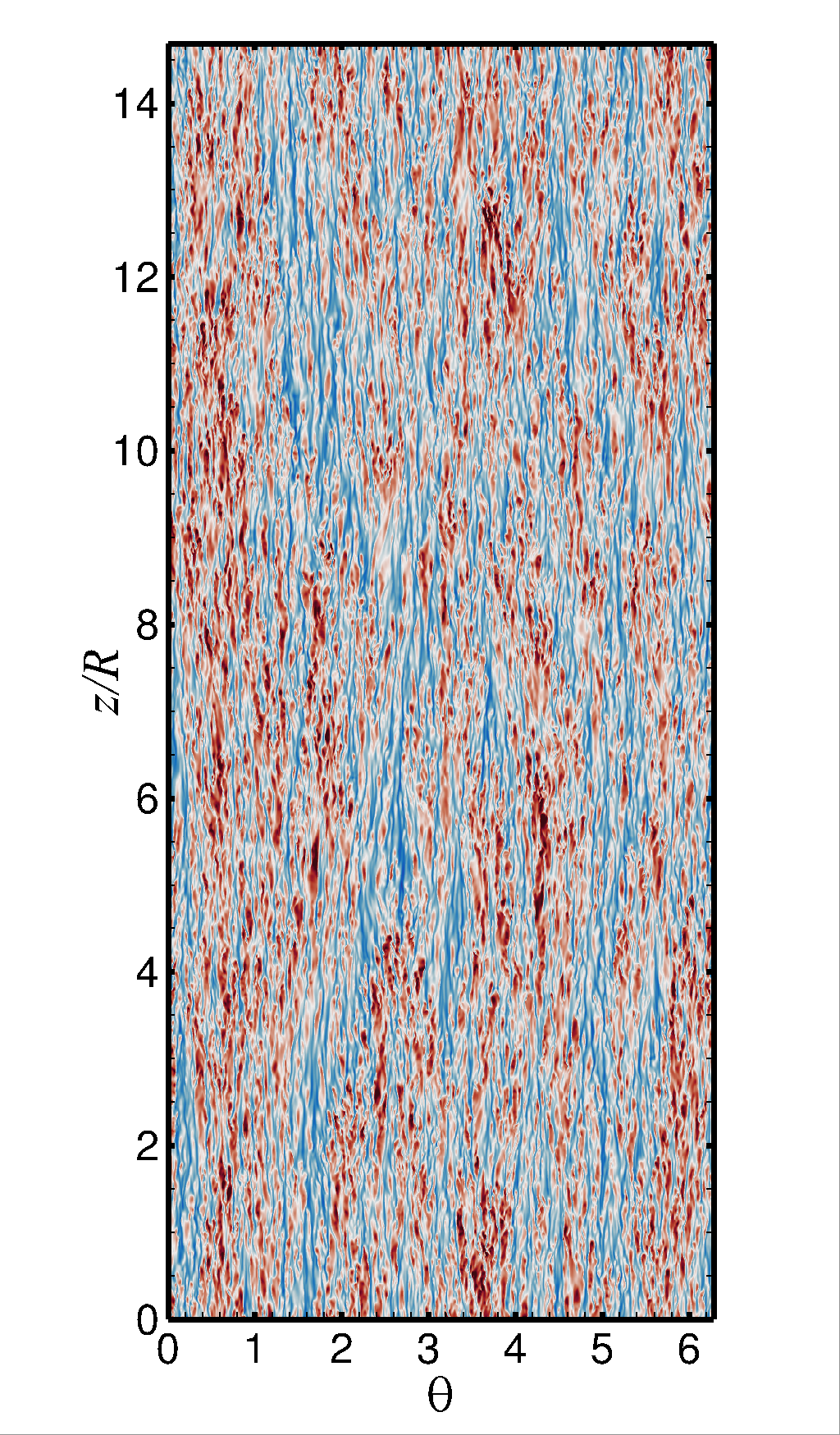} 
D~\includegraphics[width=4.0cm]{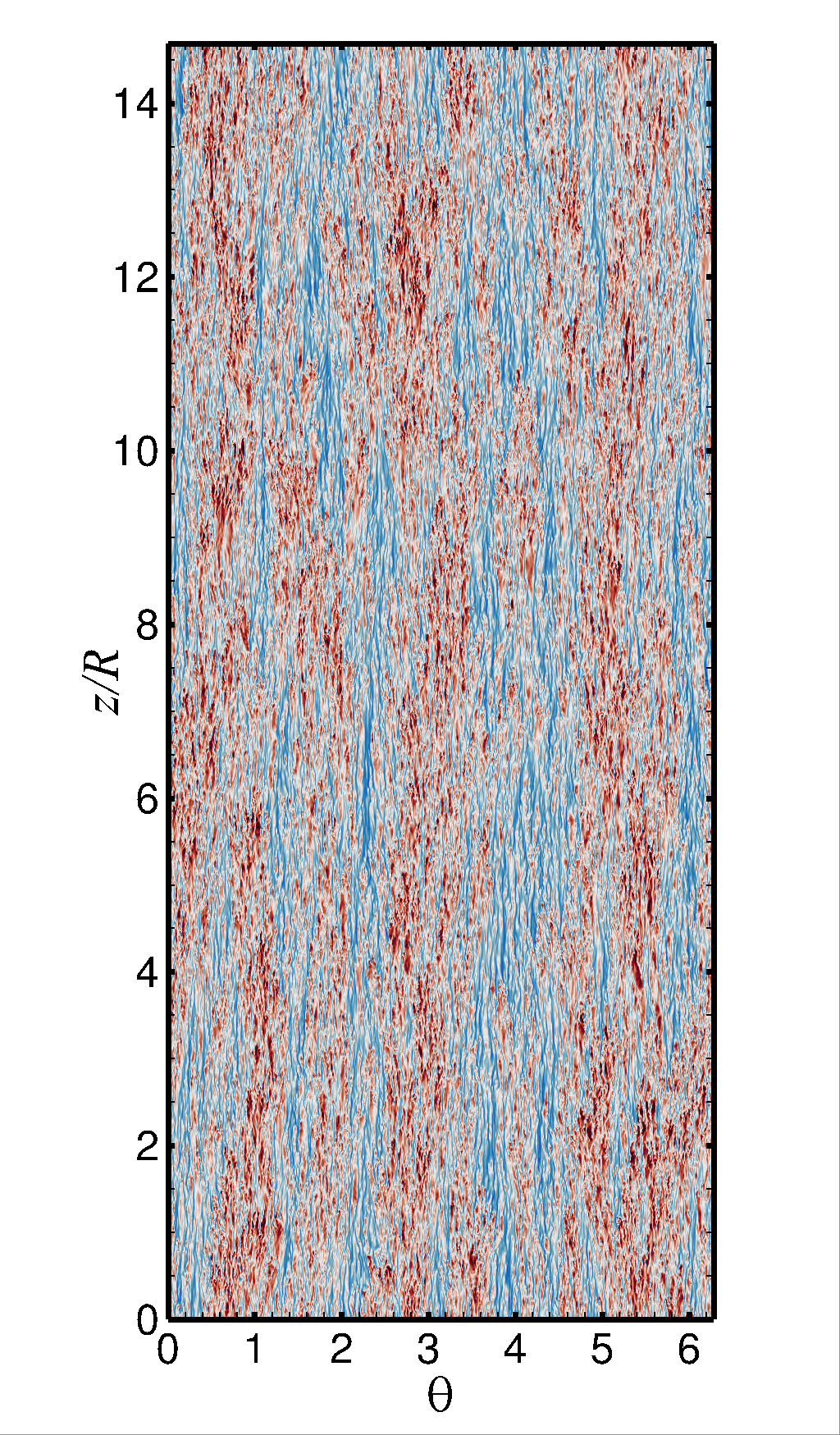} \\
E~\includegraphics[width=4.0cm]{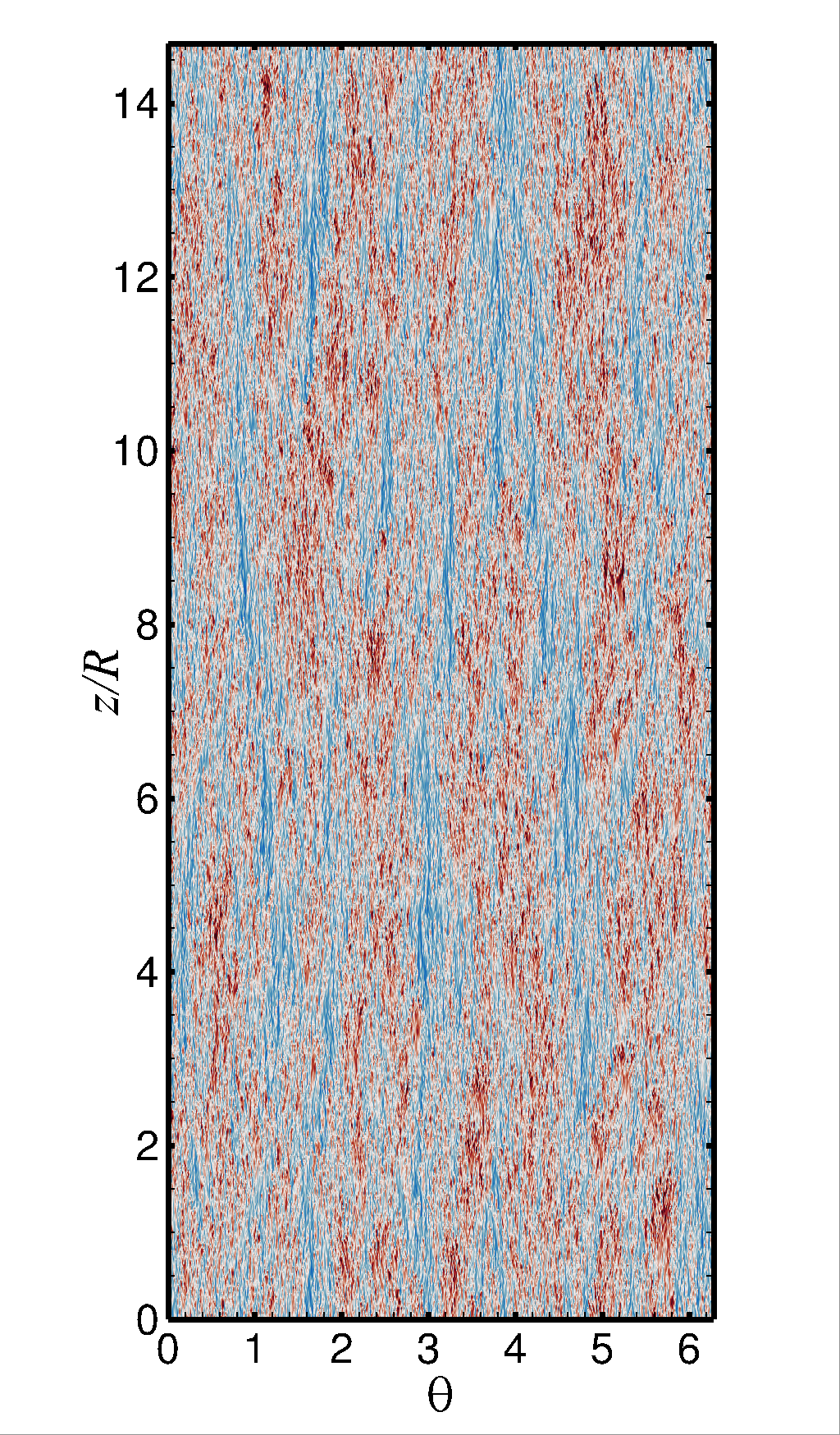} 
F~\includegraphics[width=4.0cm]{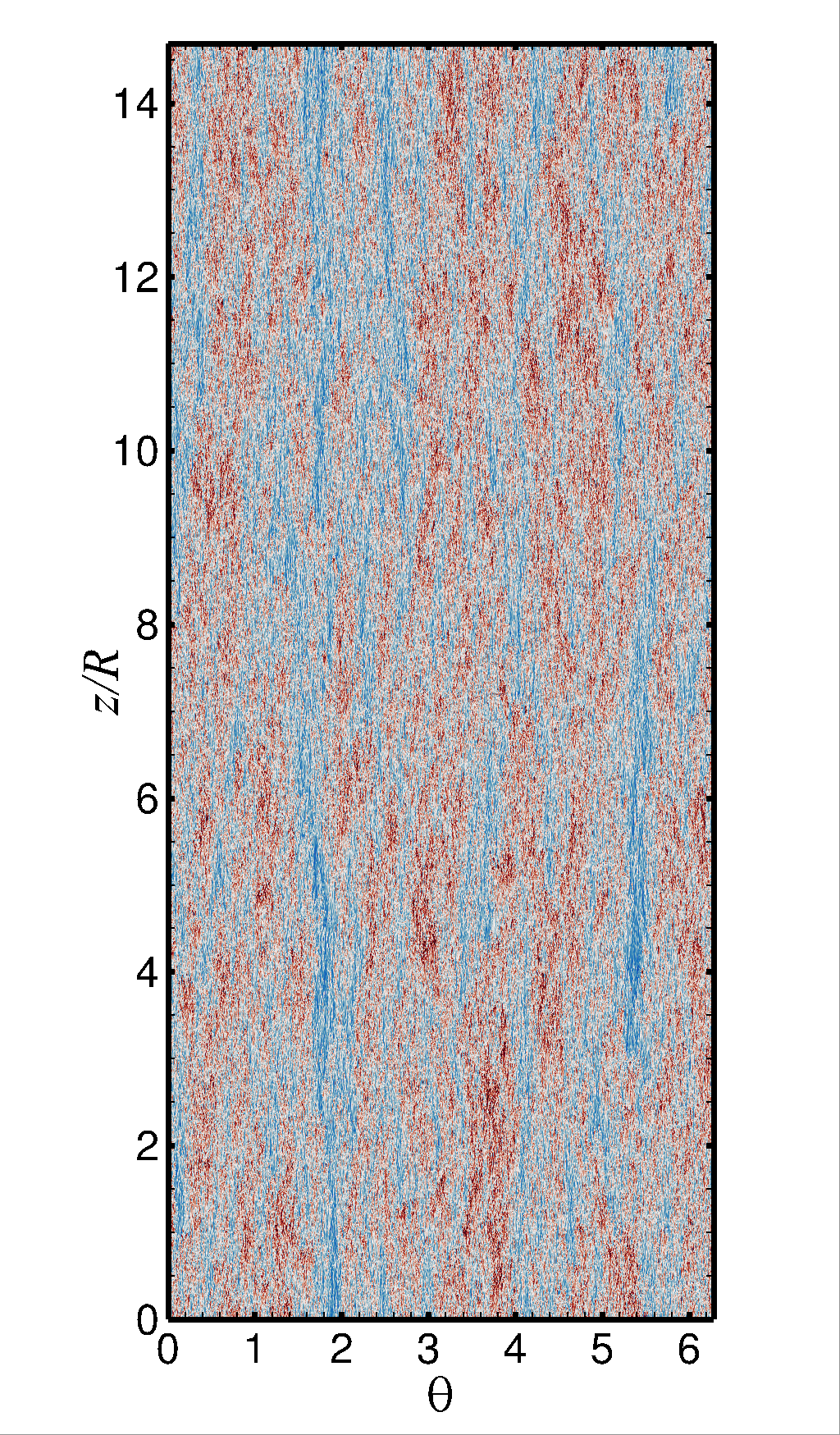}
G~\includegraphics[width=4.0cm]{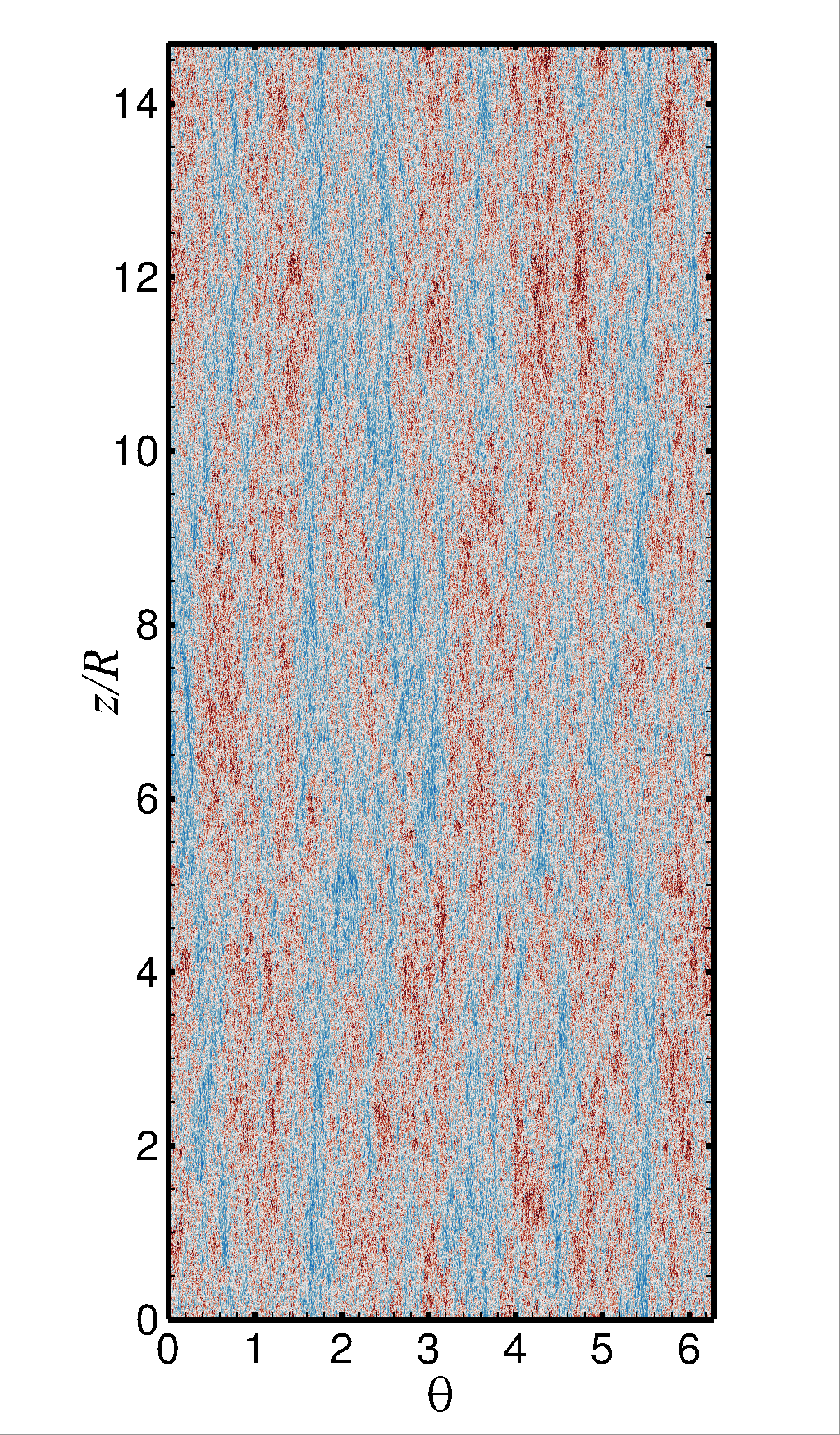} \\
\caption{Normalized streamwise velocity fluctuations ($u/\sqrt{\left< u^2 \right>}$), at $y^+ = 15$. Thirty-two contours are shown, from $-3$ to $3$, in colour scale from blue to red.}
\label{fig:streaks}
\end{figure}

The qualitative structure of flow case G (corresponding to $\Rey_{\tau} \approx 12000$) 
is not dissimilar to what observed in previous publications~\citep{pirozzoli_21}, in that the near-wall 
region is visually dominated by small-scale streaks whose size scales in inner units,
and large-scale streaks scaling on $R$. This is well portrayed in figure~\ref{fig:streaks}, which shows the 
instantaneous streamwise velocity at a distance of fifteen wall units from the wall, 
each normalized by the corresponding root-mean-square value, at various Reynolds numbers.
According to the established scenario~\citep{hutchins_07b}, the flow features a two-scale organization,
with a large number of small-scale streaks whose typical size scales in inner units, 
and superposed on those large-scale streaks being the 
footprint of 'superstructures' \citep[according to the original definition of][]{hutchins_07b}, 
whose size is instead proportional to $R$. 
Hence, whereas the former become vanishingly small as $\Rey_{\tau}$ increases, the size of the latter is visually unaffected.
Careful inspection of the figures further suggests that the superstructures seem to be most intense at the lowest Reynolds number under consideration, whereas their strength seems to slightly decline at higher $\Rey_{\tau}$.
This observation would imply that the near-wall influence of the superstructures, which are rooted in the outer layer,
slightly decreases with the Reynolds number, to an extent that we will attempt to quantify in this manuscript.

\begin{figure}
\centering
B~\includegraphics[width=6.0cm]{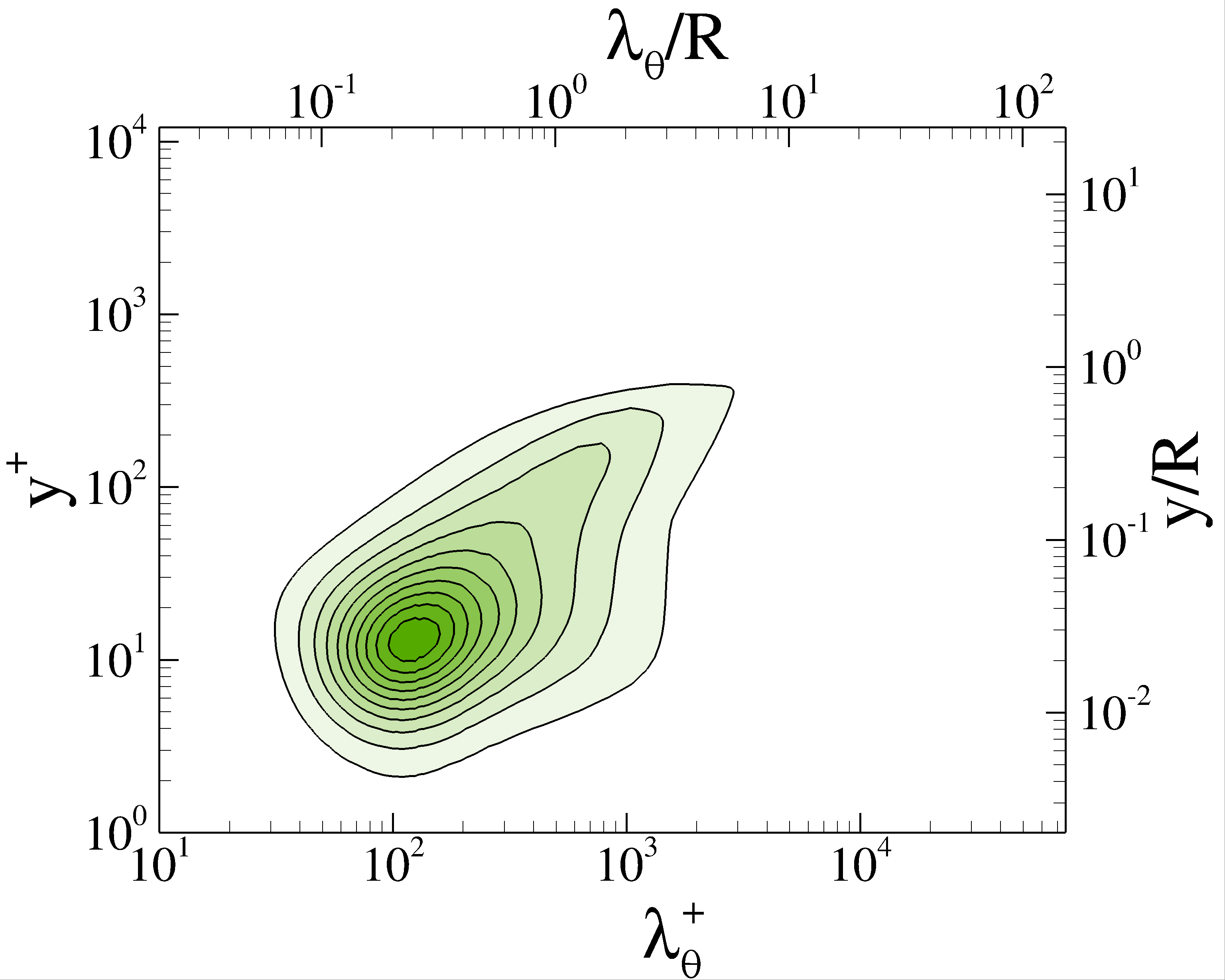}
C~\includegraphics[width=6.0cm]{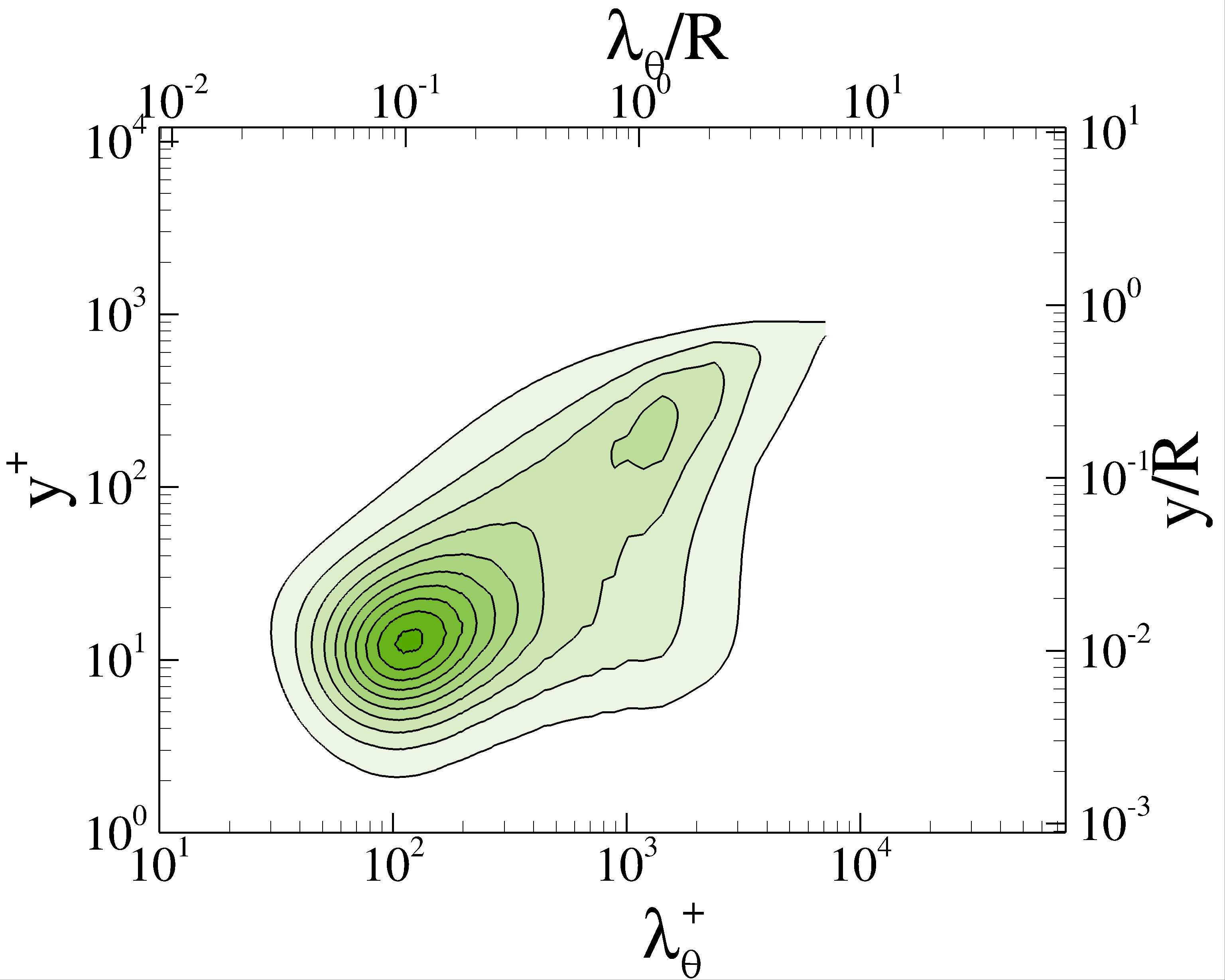} \\
D~\includegraphics[width=6.0cm]{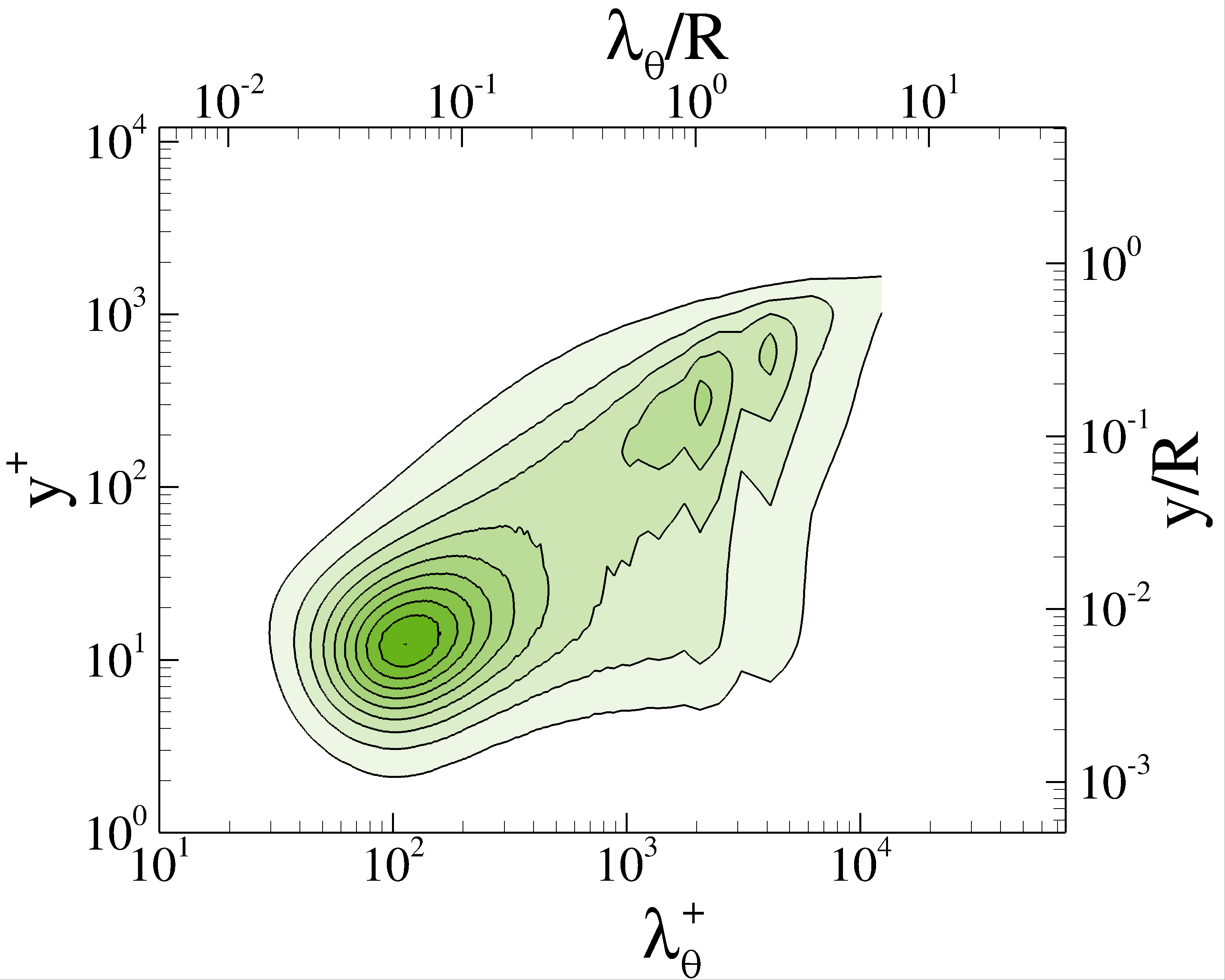}
E~\includegraphics[width=6.0cm]{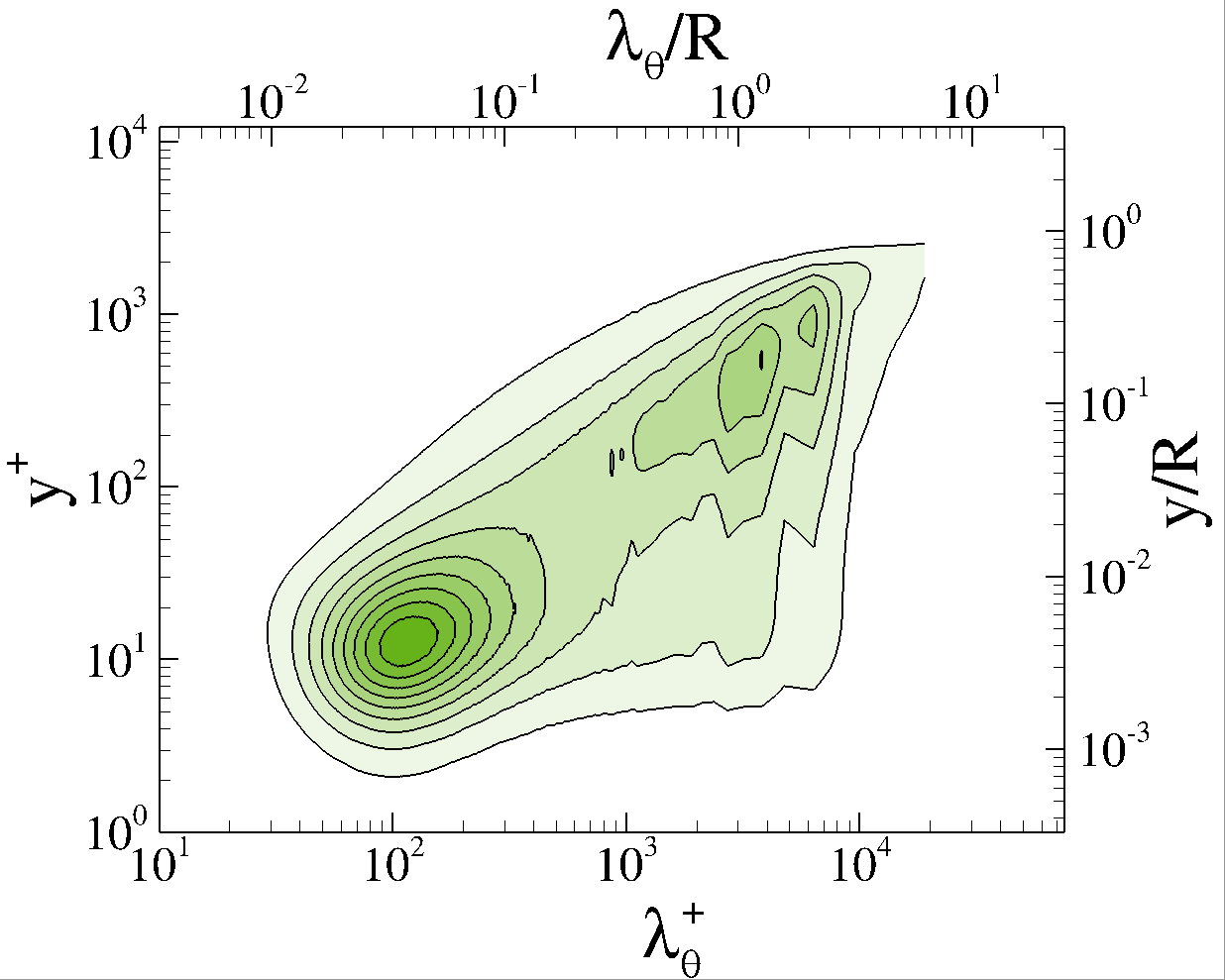} \\
F~\includegraphics[width=6.0cm]{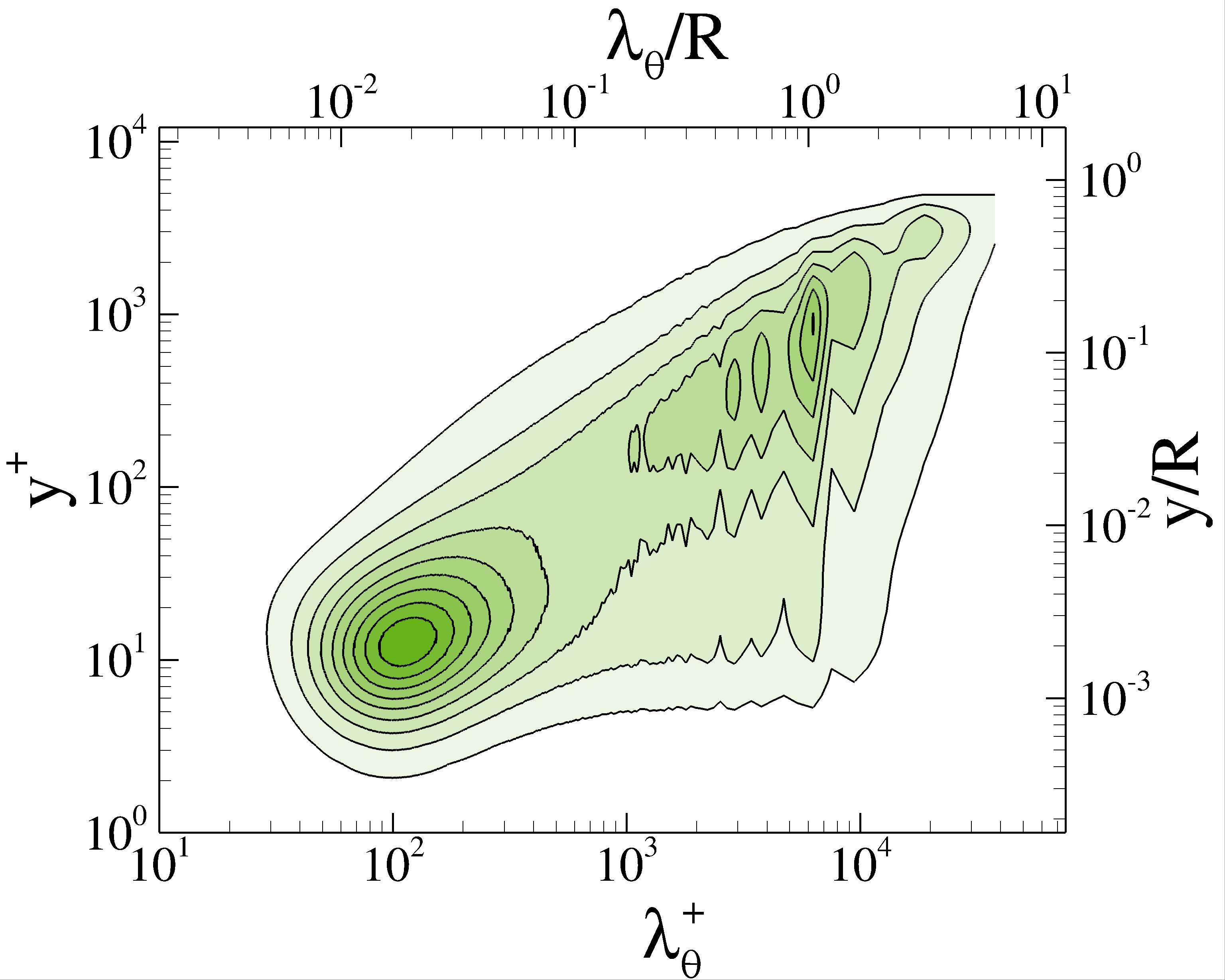}
G~\includegraphics[width=6.0cm]{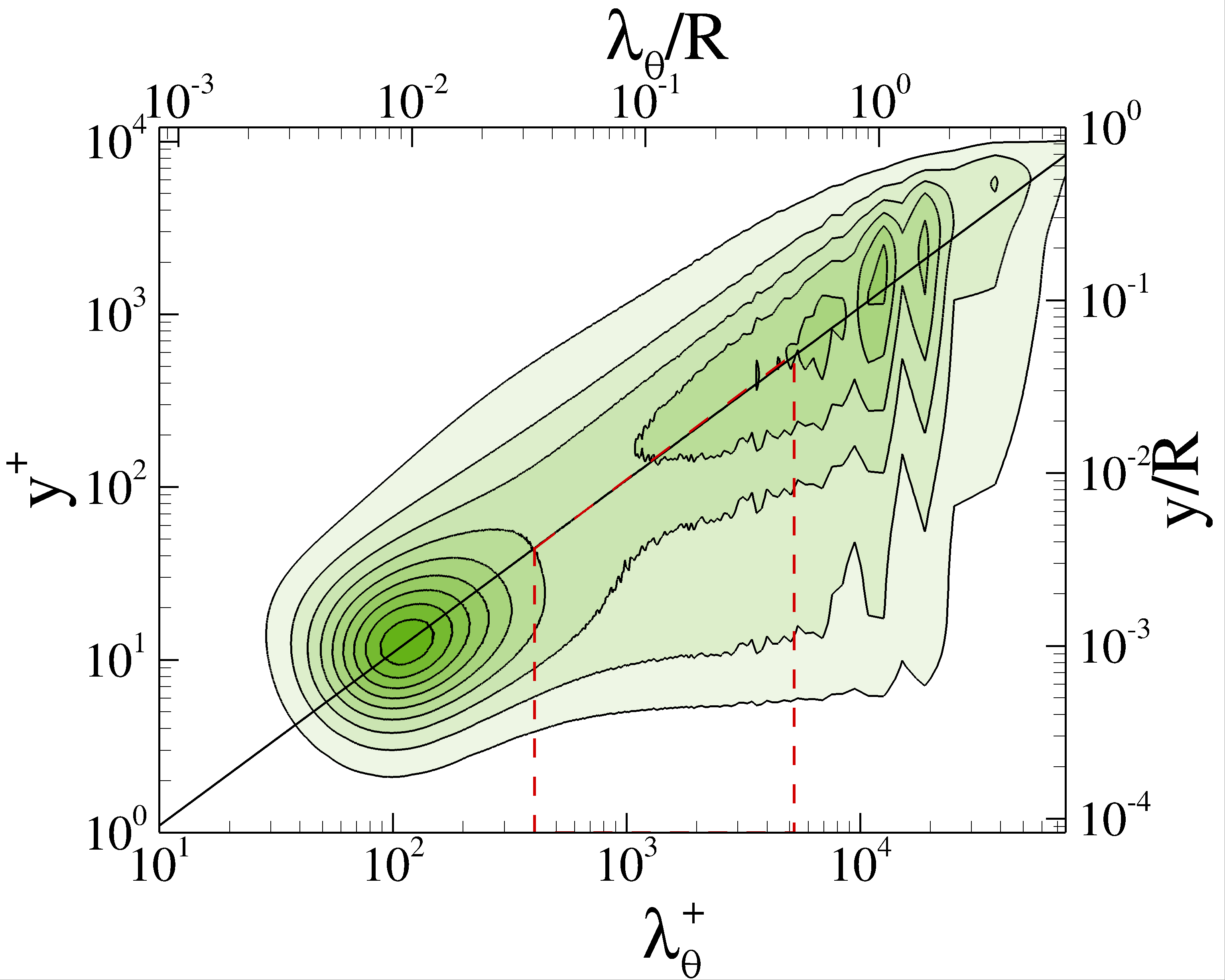} \\
     \caption{Variation of pre-multiplied spanwise spectral densities of fluctuating streamwise velocity with wall distance.
     Wall distances and wavelengths are reported both in inner units (bottom, left axes), and in outer units (top, right axes).
	In (G) the diagonal line denotes the trend $y^+ = 0.11 \lambda_{\theta}^+$, and the 
	trapezoidal region bounded by the red dashed line marks the region of near-wall influence of attached eddies.
     Contour levels from 0.36 to 3.6 are shown, in intervals of 0.36.}
\label{fig:spectra}
\end{figure}

Figure~\ref{fig:spectra} shows the spectral maps for the streamwise velocity field,
namely the spectral densities ($E_u$), presented as a function of the wall distance ($y$)
and of the spanwise wavelength ($\lambda_{\theta}$), pre-multiplied by $k_{\theta}=2 \pi / \lambda_{\theta}$.
The figure highlights near universality of the small scales of motion, with a prominent buffer-layer peak
which is universal in innewallr units, and an outer energetic site featuring
$R$-sized very large-scale motions previously observed in experiments~\citep{kim_99,hellstrom_14}, 
and which correspond to the superstructures for the present flow. 
Between the two primary locations, a band of energetic intermediate modes is observed, 
with lengths roughly proportional to their distance from the wall, 
aligning with the attached-eddy model~\citep{hwang_15}.
The discrepancy between the two sites, in terms of both physical distance and of eddy size,
increases in proportion to $\Rey_{\tau}$, reaching about two orders of magnitude in flow case G.
The figure also well clarifies that the influence of the attached eddies and of the
$O(R)$ eddies on streamwise velocity fluctuations extends down to the wall.
Indeed, based on dimensional arguments, spectral densities in the attached-eddy region are expected to
depend on $\lambda_{\theta}/y$, hence the corresponding iso-lines are expected to come in
bands parallel to the main energetic ridge, which we highlight with a diagonal line in panel G.
While this is approximately true in the region above the main ridge, the spectral iso-lines rather tend to attain a triangular
shape in the region between the spectral ridge and the wall, as a result of energy 'leakage' from the overlying eddies.
This region under the influence of wall-attached eddies is tentatively marked with dashed red lines in figure
\ref{fig:spectra}G, showing that an upper value of the wall distance exists 
past which the influence of outer eddies is not felt. 
Setting the maximum wavelength of the attached eddies to $\lambda_{\theta} = 0.4 R$,
it turns out that the maximum wall distance at which their influence is felt is 
$y_{\max}^+ \approx 0.0044 \, \Rey_{\tau}$, with $y_{\max}^+ \approx 530$ at the highest Reynolds number under scrutiny here.
This 'near-wall' region is the main subject of investigation in the present study. 

\section{The spectra of streamwise velocity}

Previous models of the velocity spectra in turbulent wall layers are mainly based on the work of \citet{perry_86}. 
The key idea is that, away from the wall, in the inviscid-dominated region, the spectra can be distinguished
in three regions: the range of dissipative eddies, scaling in inner units, the region of the $\delta$-sized eddies (here, $R$-sized), 
scaling in outer units, and a range of wall-attached eddies, for which the relevant length scale is the wall distance.
In this region, under the crucial assumption that the population density of eddies that varies inversely 
with size and hence with distance from the wall, \citet{perry_86} predicted the occurrence of a $k^{-1}$ spectral range,
where $k$ is any wall-parallel wavenumber.
Integration of the resulting spectra yields the prediction, consistent with the phenomenological theory of \citet{townsend_76}, 
that the streamwise velocity variance should decay logarithmically with the outer-scaled wall distance 
in the region of the wall layer controlled by the attached eddies.
Whereas this scenario is qualitatively confirmed in the spectral maps obtained from experiments and DNS, quantitative evidence for the $k^{-1}$ spectral range is quite scarce~\citep{vallikivi_15,baars_20}, and the original authors~\citep{perry_86} 
also observed deviations from the expected trend.
\citet{marusic_19} and \citet{baars_20} noted that limited Reynolds numbers could be a reason for failure in observing 
the expected scaling. In any case, it is not quite clear why and how the spectral scaling in the outer, 
inviscid dominated region should reflect to the near-wall region in which the peak velocity variance occurs, 
since viscous effects can be non-negligible~\citep{hwang_16}.
Arguments for why this should be the case were, nonetheless, offered by \citet{marusic_17,baars_20a}.

\begin{figure}
\centering
(a)~\includegraphics[width=6.0cm]{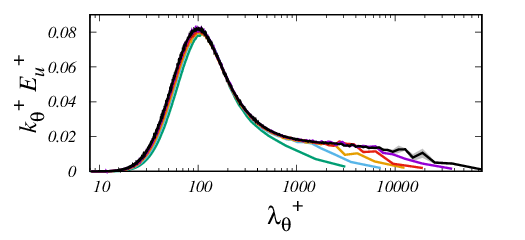}  ~(b)~\includegraphics[width=6.0cm]{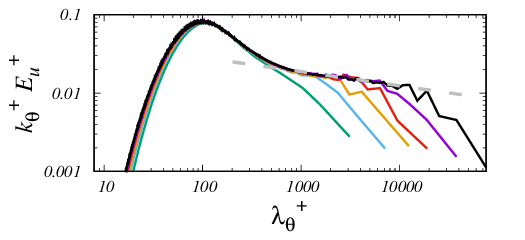} \\
(c)~\includegraphics[width=6.0cm]{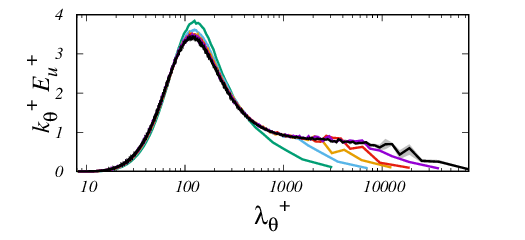} ~(d)~\includegraphics[width=6.0cm]{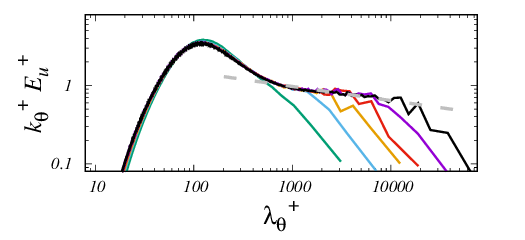} \\
(e)~\includegraphics[width=6.0cm]{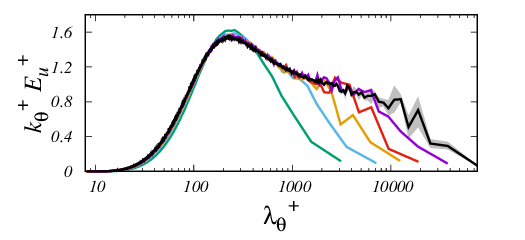} ~(f)~\includegraphics[width=6.0cm]{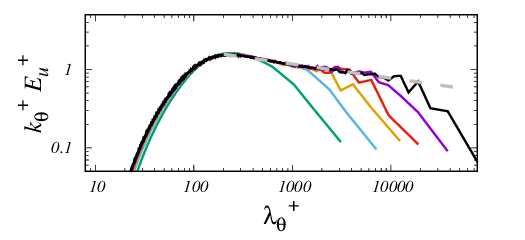} \\
(g)~\includegraphics[width=6.0cm]{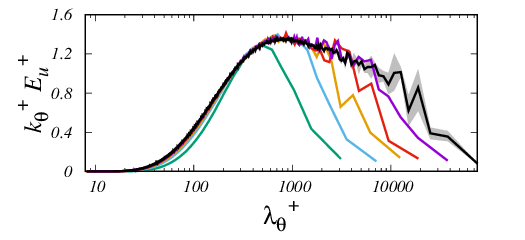}~(h)~\includegraphics[width=6.0cm]{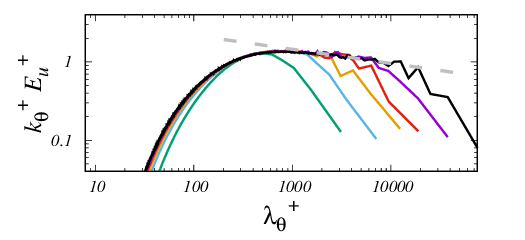} \\
(i)~\includegraphics[width=6.0cm]{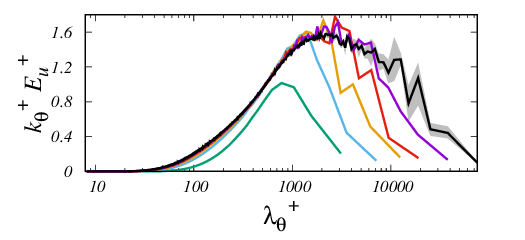}~(j)~\includegraphics[width=6.0cm]{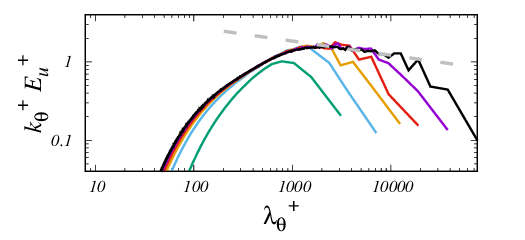} \\
(k)~\includegraphics[width=6.0cm]{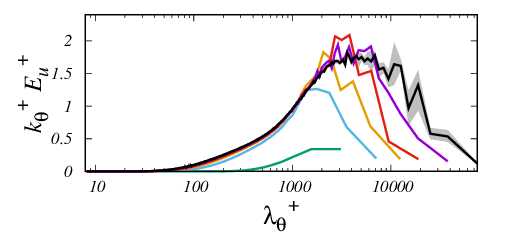}~(l)~\includegraphics[width=6.0cm]{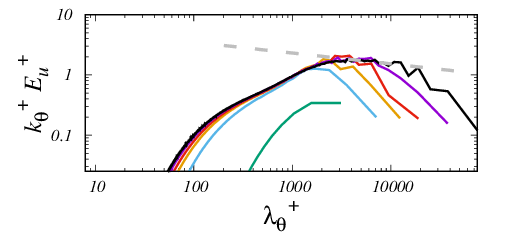} \\
\caption{Pre-multiplied spanwise spectral densities of streamwise velocity at various wall distances:
$y^+=1$ (a,b), $y^+=15$ (c,d), $y^+=50$ (e,f), $y^+=100$ (g,h), $y^+=200$ (i,j), $y^+=400$ (k,l).
A semi-log representation is used in the left-hand-side panels, and a log-log representation is used
in the right-hand-side panels. The shaded grey regions in the left-hand-side panels denotes the expected range of uncertainty for flow case G. The color codes are as in table~\ref{tab:runs}.}
\label{fig:specz_inner}
\end{figure}

In figure~\ref{fig:specz_inner} we show the spanwise spectral densities of streamwise velocity fluctuations at a number of locations at fixed $y^+ < y_{\max}^+$.
Uncertainty bars are reported in grey shades in the left-hand-side panels for flow case G, showing that 
effects of limited time convergence are mainly concentrated at scales $\lambda_{\theta} \gtrsim R$.
The figure provides clear evidence that universality of the spectra at the small scales is achieved when inner scaling is used, 
at least for $\Rey_{\tau} \gtrsim 2000$. Furthermore, this universal inner-scaled layer 
tends to become more and more extended towards longer wavelengths as $\Rey_{\tau}$ increases, until transition occurs to a $R$-scaled spectral range. 
No clear evidence for a definite $k_{\theta}^{-1}$ spectral range is observed,
which in the pre-multiplied representation would correspond to a plateau region.
Careful inspection of the velocity spectra in log-log representation 
(see the right-hand-side panels) rather suggests the existence of
a range of wavelengths with negative power-law behavior.

Standard overlap arguments as those used by \citet{millikan_38} to infer the behavior 
of the mean velocity profile in wall-bounded flows can also 
be applied to determine the plausible structure of the 
velocity spectra, by assuming that: i) the typical velocity of all eddies is the friction velocity; and
ii) the typical length of the small eddies is $\delta_v$, wheres the typical length
of the large eddies is $R$.
The transition (overlap) layer between the inner- and outer-scaled end of the 
velocity spectra should then have either the form of a logarithmic law, 
or of a power law. Based on the DNS data, the second option appears to be more appropriate, and
in that case the spectral densities in the overlap layer should behave as
\begin{equation}
	k_{\theta}^+ E_u^+ = C \left( \lambda_{\theta}^+ \right)^{-\alpha} = C \Rey_{\tau}^{-\alpha} \left( \frac{\lambda_{\theta}}R \right)^{-\alpha}, \label{eq:overlap}
\end{equation}
holding in inner and outer scaling respectively, where $\alpha$ is a possibly 
universal constant, and $C$ is a constant which in general could depend on $y^+$.
Equation~\eqref{eq:overlap} includes the $k_{\theta}^{-1}$ spectral scaling as a special case,
occurring for $\alpha=0$. In that case, both the small-scale and the large-scale end of the
spectrum should be universal.

The DNS data indeed support equation~\eqref{eq:overlap}, 
with $\alpha \approx 0.18 \pm 0.016$, as we have estimated by fitting the DNS data 
for flow case G in the range of wavelengths $1000 \le \lambda_{\theta}^+ \le 10000$. 
Notably, the power-law exponent is the same at all wall distances and all Reynolds numbers, within the numerical uncertainty. 
The power-law range is found to be widest at $y^+ = 50$.
Closer to the wall, the strong buffer-layer spectral peak
tends to mask this range towards the small wavelengths, whereas farther from the wall the 
region under influence of attached eddies tends to progressively shrink as $y^+_{\max}$ is approached.
This is in our opinion a very important observation, which has a number of implications.

\begin{figure}
\centering
(a)~\includegraphics[width=6.0cm]{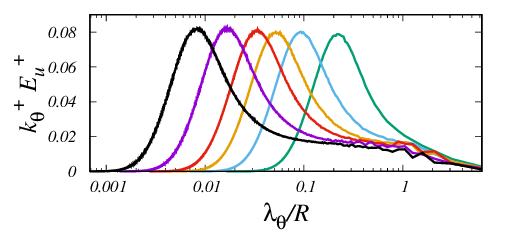}  
(b)~\includegraphics[width=6.0cm]{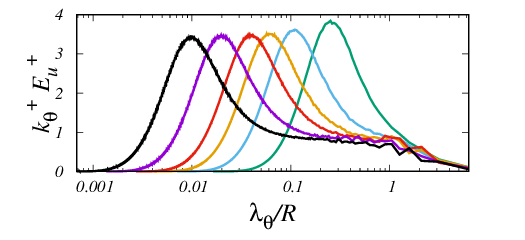} 
(c)~\includegraphics[width=6.0cm]{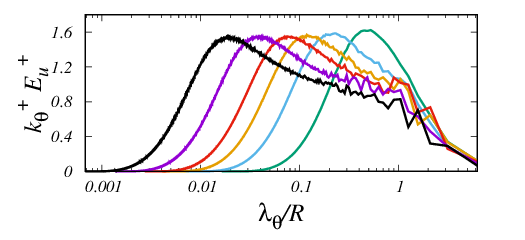} 
(d)~\includegraphics[width=6.0cm]{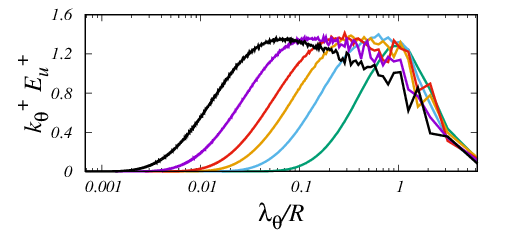}
(e)~\includegraphics[width=6.0cm]{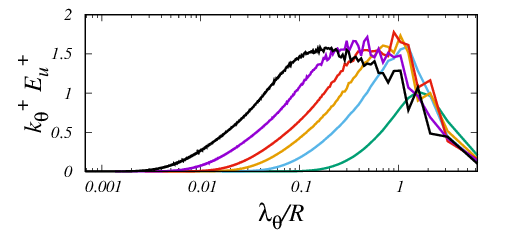}
(f)~\includegraphics[width=6.0cm]{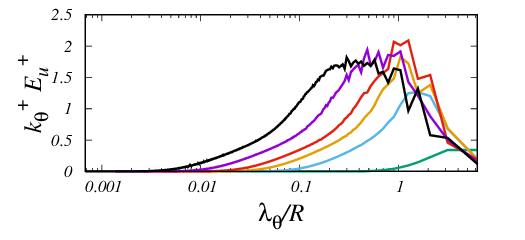}
\caption{Pre-multiplied spanwise spectral densities of streamwise velocity at various wall distances, reported in outer scaling at $y^+=1$ (a), $y^+=15$ (b), $y^+=50$ (c), $y^+=100$ (d), $y^+=200$ (e), $y^+=400$ (f).
The color codes are as in table~\ref{tab:runs}.}
\label{fig:specz_out}
\end{figure}

\begin{figure}
\centering
(a)~\includegraphics[width=6.0cm]{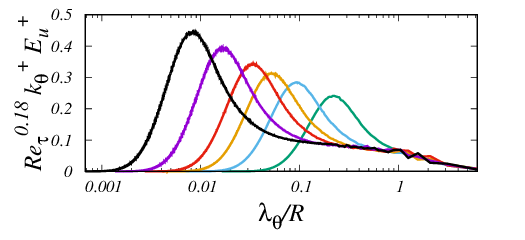}  
(b)~\includegraphics[width=6.0cm]{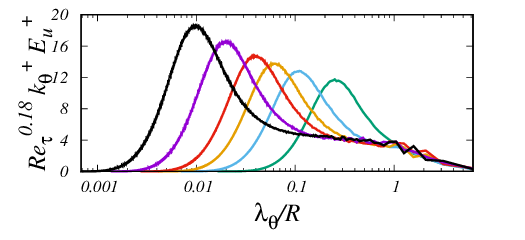} 
(c)~\includegraphics[width=6.0cm]{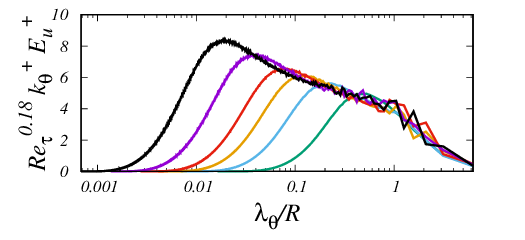} 
(d)~\includegraphics[width=6.0cm]{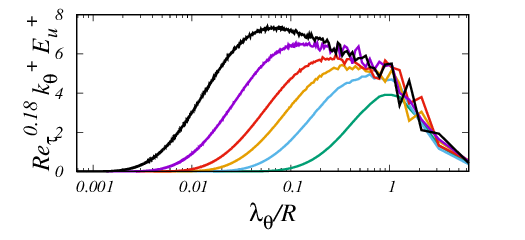}
(e)~\includegraphics[width=6.0cm]{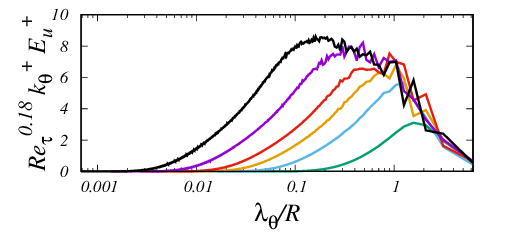}
(f)~\includegraphics[width=6.0cm]{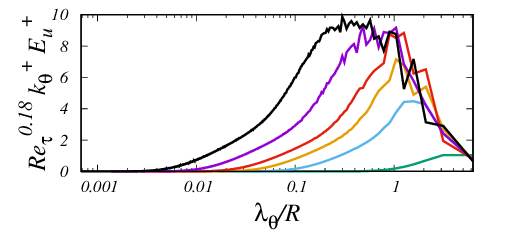}
\caption{Pre-multiplied spanwise spectral densities of streamwise velocity at various wall distances, reported in outer scaling and compensated by $\Rey_{\tau}^{0.18}$: $y^+=1$ (a), $y^+=15$ (b), $y^+=50$ (c), $y^+=100$ (d), $y^+=200$ (e), $y^+=400$ (f).
The color codes are as in table~\ref{tab:runs}.}
\label{fig:specz_out_comp}
\end{figure}

To test universality of the large-scale end of the spectra, in figure~\ref{fig:specz_out}, 
the spectral densities are shown as a function of the outer-scaled wavelength.
Universality is clearly not as good as seen for the small wavelengths in figure~\ref{fig:specz_inner}, 
with the energy associated with the large scales of motion becoming slightly but consistently less at higher Reynolds number, for given $\lambda_{\theta}/R$.
In figure~\ref{fig:specz_out_comp} we then show the outer-scaled spectra compensated by $\Rey_{\tau}^{\alpha}$, as suggested from equation~\eqref{eq:overlap}.
The figure shows that, with good accuracy, universality at the large-scale end of the spectrum 
is recovered in a wide range of scales, up to a short-wavelength limit which 
is shifting to the left as $\Rey_{\tau}$ increases.

Overall, convincing evidence exists that the small-scale end of the velocity spectra
is universal, whereas the large-scale end is not, with an overlap
layer connecting the two ends which features
a negative power-law behavior with exponent $\alpha \approx 0.18$.
The same behaviour is also traced in spectra from plane channel flow DNS~\citep{lee_15},
as shown in appendix~\ref{sec:channel}, which corroborates the generality of our findings.

\section{The streamwise velocity variance}

The information derived from the analysis of the velocity spectra can be distilled to infer the behaviour of the velocity variance,
taking inspiration from~\citet{hwang_23}. 
Let $\lambda_s$ and $\lambda_{\ell}$ be, respectively, indicative lower and upper limits for
the observed overlap spectral range, the velocity variance can be expressed as
\begin{equation}
	\left< u^2 \right>^+ = 
	\underbrace{\int_{0}^{\lambda_s^+} k_{\theta}^+ E_{u,s}^+ \diff \log \lambda_{\theta}^+}_{<u^2>^+_s} +
	\underbrace{\int_{\lambda_s^+}^{\lambda_{\ell}^+} k_{\theta}^+ E_{u,o}^+ \diff \log \lambda_{\theta}^+}_{<u^2>^+_o} +
	\underbrace{\int_{\lambda_{\ell}^+}^{\infty} k_{\theta}^+ E_{u,\ell}^+ \diff \log \lambda_{\theta}^+}_{<u^2>^+_{\ell}}, \label{eq:split}
\end{equation}
where the subscripts $s$, $\ell$ and $o$ denote, respectively, the contributions of the smallest scales, 
the largest scales, and the intermediate, overlap-layer scales.
Although the precise values of the limits in \eqref{eq:split} are not important, it is crucial 
that the lower limit scales in inner units, hence $\lambda_{s}^+ = \mathrm{const}$, 
and that the upper limit scales in outer units, hence $\lambda_{\ell}/R = \mathrm{const}$. 
Based on the evidence previously given that the smallest scales tend to be universal across the $\Rey_{\tau}$ range,
the associated contribution to the velocity variance is also expected to be asymptotically constant, namely
\begin{equation}
	\left< u^2 \right>^+_s = A_s (y^+). \label{eq:urms2_s}
\end{equation}
Since the upper end of the overlap layer should scale in outer units, we further have 
$\lambda_{\ell}^+ = \Rey_{\tau} \lambda_{\ell}/R \sim \Rey_{\tau}$.
Based on the matching condition~\eqref{eq:overlap}, and on inspection of 
figure~\ref{fig:specz_out_comp}, the contribution from the large scales 
is then expected to vary as
\begin{equation}
	{\left< u^2 \right>^+_{\ell}} = B_{\ell} (y^+) \, \Rey_{\tau}^{-\alpha} . \label{eq:urms2_l}
\end{equation}
This is an especially significant formula implying that imprinting effects imparted by the largest scales of motion (superstructures)
on the near-wall region should actually decrease as $\Rey_{\tau}$ increases,
in line with observations made by \citet{hwang_23}. 
Last, the contribution of the overlap layer can be evaluated by integrating 
the power-law spectrum given in equation~\eqref{eq:overlap}, thus obtaining 
\begin{equation}
	{\left< u^2 \right>^+_o} = \frac{C(y^+)}{\alpha} \left[ \left( {\lambda_s^+}\right)^{-\alpha} - \left(\frac{\lambda_{\ell}}{R}\right)^{-\alpha} \Rey_{\tau}^{-\alpha} \right] = A_o(y^+) - B_o(y^+) \, \Rey_{\tau}^{-\alpha}. \label{eq:urms2_o}
\end{equation}

\begin{figure}
\centerline{
(a)
\includegraphics[width=7.0cm]{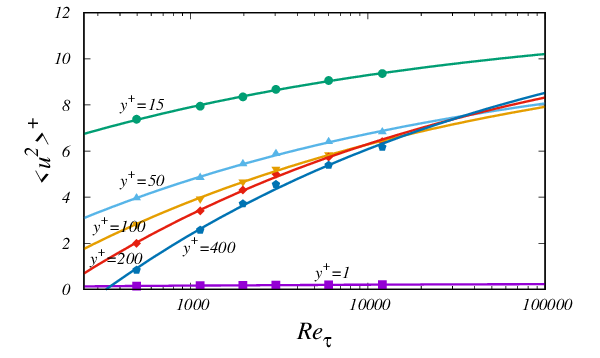}  
(b)
\includegraphics[width=7.0cm]{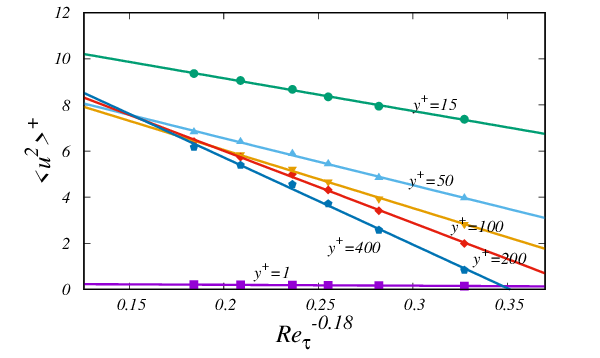}  
}
\caption{Streamwise velocity variances (symbols) as a function of $\Rey_{\tau}$ (a) and as a function of $\Rey_{\tau}^{-0.18}$ (b), at various off-wall positions, and corresponding fits, according to equation~\eqref{eq:urms2}, with coefficients given in table~\ref{tab:urms2_fit}.}
\label{fig:urms2_fit}
\end{figure}

The most important inference of the present analysis is that the overall velocity variance should then vary as 
\begin{equation}
	\left< u^2 \right>^+ = \underbrace{\left( A_s(y^+) + A_o (y^+) \right)}_{A(y^+)} - \underbrace{\left(B_o(y^+) - B_{\ell}(y^+)\right)}_{B(y^+)} \Rey_{\tau}^{-\alpha} , \label{eq:urms2}
\end{equation}
where the functions $A$ and $B$ do not depend explicitly on the Reynolds number, 
given the assumptions made to define $\lambda_s$ and $\lambda_{\ell}$.
Interestingly, this formula is formally identical to the asymptotic expansion suggested by
\citet{monkewitz_22}, in which $\Rey_{\tau}^{-\alpha}$ has the role of gauge function.
This formula predicts that, at any fixed $y^+$, the velocity variance should increase 
asymptoting to a finite limit as $\Rey_{\tau}$ increases, thus restoring strict wall scaling. 
Values for the asymptotic constants $A$, $B$ determined from fitting the DNS data at representative
off-wall locations are reported in table~\ref{tab:urms2_fit}, 
along with the corresponding standard deviations.
The resulting distributions of the streamwise velocity variances as a function 
of $\Rey_{\tau}$ are shown in figure~\ref{fig:urms2_fit}.
The quality of the fits is quite good, although it tends to
deteriorate farther from the wall, since the power-law spectral range becomes narrow,
making extrapolations not fully reliable.
Most notably, figure~\ref{fig:urms2_fit} shows that, whereas
the defect power law could well be mistaken with logarithmic growth at $y^+=15$,
the trends at positions farther from the wall are distinctly different than logarithmic.
This difference is due to the large contribution conveyed by the smallest eddies 
close to the wall (recalling figure~\ref{fig:specz_inner}), which tends to overshadow
Reynolds number variations associated with the larger eddies. The reduction of the energetic 
peak associated with the near-wall streaks occurring farther from the wall makes the actual Reynolds 
number dependence manifest.
Anyhow, the trend towards the asymptotic limit is quite slow, and even at extreme 
Reynolds numbers as $\Rey_{\tau}=10^6$, the predicted difference between the velocity variance at $y^+=15$
and its asymptotic value would still be about $8\%$.

\begin{table}
 \centering
 \begin{tabular*}{1.\textwidth}{@{\extracolsep{\fill}}lcc}
	 Station & $A(y^+)$ & $B(y^+)$ \\
 \hline
	 $y^+=1$   & $0.281 \pm 0.00257 (0.915\%)$ & $0.396 \pm 0.0101 (2.56\%)$ \\
	 $y^+=15$  & $12.0  \pm 0.0808  (0.672\%)$ & $14.2  \pm 0.319  (2.25\%)$ \\
	 $y^+=50$  & $10.6  \pm 0.105   (0.985\%)$ & $20.4  \pm 0.413  (2.03\%)$ \\	
	 $y^+=100$ & $11.1  \pm 0.133   (1.20 \%)$ & $25.2  \pm 0.526  (2.09\%)$ \\
	 $y^+=200$ & $12.3  \pm 0.185   (1.51 \%)$ & $31.2  \pm 0.729  (1.51\%)$ \\
	 $y^+=400$ & $13.3  \pm 0.307   (2.31 \%)$ & $37.8  \pm 1.21   (3.21\%)$ \\
 \hline
 \end{tabular*}
	\caption{Fitting parameters to use in equation~\eqref{eq:urms2}, based on DNS data fitting, at several off-wall positions, with accompanying asymptotic standard errors ($\alpha = 0.18$ is assumed).}
\label{tab:urms2_fit}
\end{table}

\begin{figure}
\centering
\includegraphics[width=10.0cm]{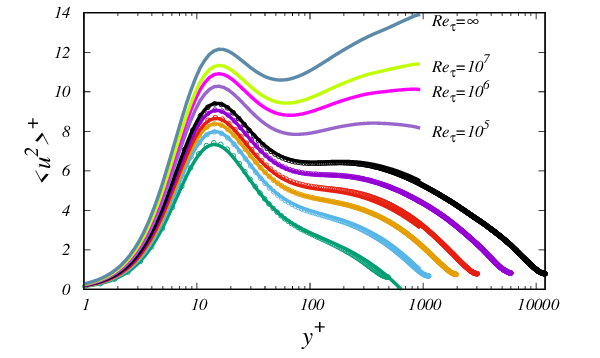}  
\caption{Predicted distributions of streamwise velocity variances at various $\Rey_{\tau}$, according to equation~\eqref{eq:urms2}, assuming $\alpha = 0.18$. The symbols denote the DNS data used to determine the fit coefficients $A(y^+)$, $B(y^+)$ (see table~\ref{tab:runs} for the color codes).}
\label{fig:urms2_extr}
\end{figure}

The extrapolated distributions of the streamwise velocity variance as a function of the wall distance
are shown in figure~\ref{fig:urms2_extr}, for various $\Rey_{\tau}$. While confirming that the 
predictive formula~\eqref{eq:urms2} covers well the range of Reynolds numbers for which the model is trained, 
the figure also shows extrapolations beyond that range.
Regarding the buffer-layer peak, it is predicted to remain more or less at the same position in inner units, 
and its maximum amplitude at infinite Reynolds number is predicted to be approximately 12.1.
Taking into consideration the uncertainty associated with the parameter $\alpha$, 
the buffer-layer velocity variance peak is estimated to fall within the range of 11.9 to 12.5. 
It is worth mentioning that these values are somewhat higher than the value of 11.5 derived 
from the $\Rey_{\tau}^{-0.25}$ defect law \citep{chen_21}, while \citet{monkewitz_22} 
obtained a value of 11.3 through an inner asymptotic expansion informed by DNS data. 
Farther from the wall, the distribution tends to form a shoulder as $\Rey_{\tau}$ increases, with the eventual
onset of a clear 'outer peak'. This outer peak is barely visible at the DNS-accessible Reynolds numbers, 
but based on the extrapolated data it should attain a higher value than the inner peak at extreme $\Rey_{\tau}$. 

\begin{figure}
 \begin{center}	
 (a)~\includegraphics[width=7.0cm]{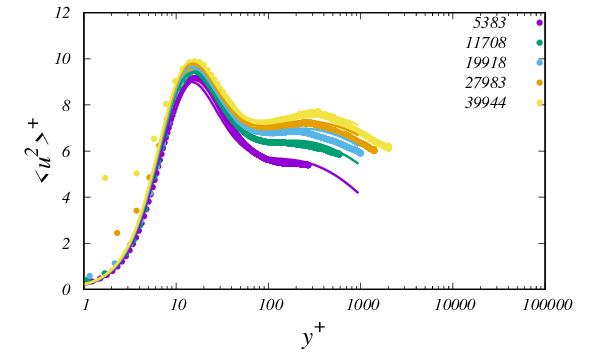} \\
 (c)~\includegraphics[width=7.0cm]{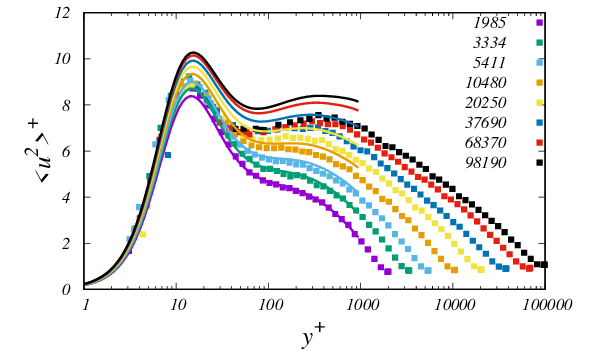}
 (c)~\includegraphics[width=7.0cm]{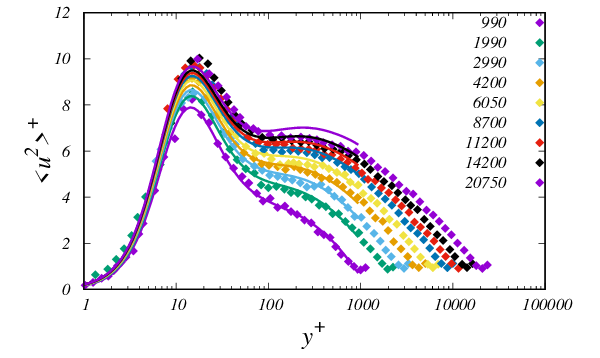}  
 \end{center}
\caption{Comparison of streamwise velocity variance distributions predicted from equation~\eqref{eq:urms2} (solid lines), with experimental measurements taken in the SuperPipe facility~\citep[(b)][]{hultmark_12}, in the Ciclope facility~\citep[(c)][]{willert_17}, and in the Hi-Reff facility~\citep[(d)][]{ono_23}, at matching values of $\Rey_{\tau}$, as given in the respective legends.}
\label{fig:urms2_comp}
\end{figure}

The primary inquiry revolves around whether the extrapolated distributions agree with experimental measurements. 
To address this question, in figure~\ref{fig:urms2_comp} we present experimental data collected from various 
facilities using different measurement techniques. Specifically, particle-image velocimetry (PIV) measurements 
from the CICLoPE facility~\citep{willert_17}, hot-wire anemometry (HWA) measurements conducted with nanoscale 
thermal anemometry probes in Princeton's Superpipe facility~\citep{hultmark_12}, 
and laser Doppler velocimetry (LDV) measurements carried out in the Hi-Reff facility in Japan~\citep{ono_23} are included.
The comparison reveals highly favorable agreement with all measurements at 'low' Reynolds numbers, 
approximately $\Rey_{\tau} \lesssim 5000$. However, notable discrepancies arise at higher Reynolds numbers, 
varying considerably depending on the set of measurements. Specifically, while alignment remains nearly 
perfect with the PIV measurements in the CICLoPE facility, the DNS-based extrapolations yield values 
significantly higher than those obtained from the SuperPipe data.
A similar trend of over-prediction of the experimental data is also noticeable, 
albeit to a lesser degree, when comparing with the LDV measurements in Hi-Reff.
In principle, these differences could arise from shortcomings in the DNS and/or 
its application for extrapolations, but they are more likely associated with 
issues in the experimental data. Indeed, the three sets of measurements utilized 
here for reference demonstrate notable discrepancies among each other, suggesting 
possible filtering effects, particularly for HWA and LDV measurements, 
whereas PIV measurements might be less affected. 

The current analysis also has direct implications for the dissipation rate of the streamwise velocity variance, $\epsilon_{11} = \nu \left< \left| \nabla u \right|^2 \right>$.
As noted by \citet{chen_21}, at the wall this quantity equals the viscous diffusion term,
which in inner units is $\diff^2 \left< u^2 \right>^+ / \diff {y^+}^2$. The latter quantity (not shown) is virtually indistinguishable from 
$\left< u^2 \right>^+(y^+=1)$, since $\diff \left< u^2 \right>^+ / \diff y^+ = 0$ at the wall. 
The inferred trend of $\left< u^2 \right>^+(y^+=1)$ (see table~\ref{tab:urms2_fit}) 
then implies that the wall dissipation rate should asymptote 
to a value of about $0.28$, with uncertainty of $\pm 0.01$ on account 
of variability of $\alpha$. Although a bit higher than the upper bound of $0.25$ advocated by \citet{chen_21},
and of the value of $0.26$ predicted by \citet{monkewitz_15} for zero-pressure-gradient boundary layers, this result suggests that wall dissipation
retains the same order of magnitude as the maximum production in the buffer layer. 

\section{Discussion and Conclusions}

The current analysis, informed from spectral DNS data of turbulent pipe flow 
up to $\Rey_{\tau} \approx 12000$, raises several significant points. Firstly, 
it offers quantitative arguments, alongside physical intuition, suggesting that 
strict wall scaling should be restored in the limit of infinite Reynolds numbers. 
This assertion aligns with propositions put forth in recent literature \citep{chen_21,monkewitz_22}, 
albeit founded on entirely different arguments.
In particular, consistent with the assumptions made by \citet{chen_21}, 
it appears that the appropriate asymptotic behaviour towards the infinite-Reynolds-number 
limit follows a defect power law. However, the exponent of this negative power law appears 
to deviate somewhat from the $\alpha=0.25$ advocated in those studies, being instead closer to $\alpha = 0.18$, 
with an uncertainty of approximately $\pm 10\%$, as revealed by the analysis of the velocity spectra. 
This discrepancy might stem from the inaccurate assumption made by those authors that the wall dissipation 
rate should asymptotically approach $0.25$ not to exceed the maximum production rate, 
although this assertion cannot be rigorously justified on mathematical grounds. 

The current analysis fits well within the framework of the classical attached eddy model.
Indeed, the spectrograms reported in figure~\ref{fig:spectra} bear the clear signature of a hierarchy of wall-attached eddies.
However, in contrast to the classical formulation of the AEM,
we observe that the near-wall signature of these attached eddies becomes fainter 
as their centre moves away from the wall. This phenomenon is reflected in a spectral decay 
that is shallower than the traditionally accepted $k^{-1}$ spectrum derived from inviscid arguments.
A secondary implication is that the imprinting of superstructures (here, $R$-sized eddies) 
on the wall diminishes as $\Rey_{\tau}$ increases. However, this observation does not imply 
that inner/outer layer interactions are negligible. Rather, the slow decay rate implies 
that the influence of outer-layer eddies remains substantial at any reasonably high Reynolds number.

\begin{figure}
\centerline{
(a)
\includegraphics[width=7.0cm]{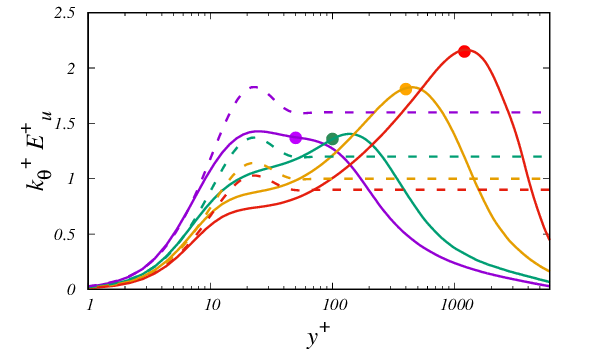}  
(b)
\includegraphics[width=7.0cm]{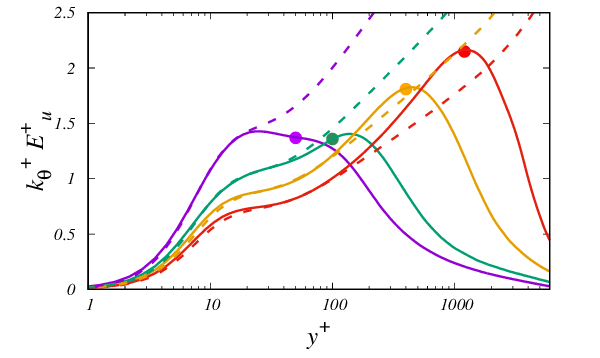}  
}
\caption{Flow case G: pre-multiplied spectral density of streamwise velocity as a function of wall distance, corresponding to various wavelengths: 
$\lambda_{\theta}^+=455$  ($y_s^+=50$) (purple),
$\lambda_{\theta}^+=909$  ($y_s^+=100$) (green),
$\lambda_{\theta}^+=3636$ ($y_s^+=400$) (orange),
$\lambda_{\theta}^+=10970$ ($y_s^+=1205$) (red).
The bullets denote the wall distance of the corresponding eddy centers, see figure~\ref{fig:spectra}.
The dashed lines in (a) denote predictions of equation~\eqref{eq:stokes} with $\Delta^+=10$, and in (b) predictions of equation~\eqref{eq:bradshaw}.
}
\label{fig:stokes}
\end{figure}

The crucial question revolves around the reason for the spectral slope being less steep than $k^{-1}$. 
A plausible explanation is the presence of viscous effects, which impede the flow in the near-wall region. 
While the AEM treats inactive motions (whose size significantly exceeds the distance from the wall) 
as satisfying a slip condition at the wall, in reality, they do not. 
In this regard, the analysis conducted by \citet{spalart_88} holds considerable merit. 
He argued that inactive eddies influence the near-wall region by inducing periodic, 
low-frequency sloshing motions, leading to the formation of Stokes layers of either laminar or turbulent nature.
In the first case, analytical solution of the Navier-Stokes equations with time-periodic change of 
the free-stream velocity would yield the following form of the streamwise velocity variance~\citep{schlichting_00}
\begin{equation}
\left< u^2 \right>^+ \sim 1 + e^{-2 y / \Delta} - 2 \cos(y/\Delta) e^{-y/\Delta},
\label{eq:stokes}
\end{equation}
with $\Delta = (2 \nu / \omega)^{1/2}$ the thickness of the Stokes layer, and 
$\omega$ the oscillation frequency.
\citet{bradshaw_67} proposed a model for turbulent Stokes layers based on the idea that
the primary effect of low-frequency inactive motions is a time-periodic change of the wall shear stress,
namely $\tau_w = \overline{\tau}_w \left( 1 + A \cos (\omega t) \right)$, with $A << 1$.
Assuming next that the law-of-the-wall has time to adapt to low-frequency modulation of the wall
shear stress results in the following prediction for the velocity variance associated with
inactive motions,
\begin{equation}
\left< u^2 \right>^+ \sim \left( U^+ + y^+ \frac{\diff U^+}{\diff y^+} \right)^2.
\label{eq:bradshaw}
\end{equation}
These concepts were put to the test by~\citet{spalart_88}, 
who compared the predictions in equations~\eqref{eq:stokes} and~\eqref{eq:bradshaw} 
with the difference in velocity variances at two modest Reynolds numbers, which he 
used as a proxy for the contribution of inactive motions. Here, both approaches are 
examined by plotting profiles of $E_u^+(\lambda_{\theta}^+, y^+)$ for various values of $\lambda_{\theta}^+$, 
which in the AEM interpretation characterize the intensity of streamwise velocity fluctuations induced by 
attached eddies centered at a wall distance $y_s^+ \approx 0.11 \lambda_{\theta}^+$ (refer to Figure \ref{fig:spectra}).

Figure~\ref{fig:stokes}(a) presents a comparison of the spectral density profiles 
with the prediction given by equation~\eqref{eq:stokes}, where we have arbitrarily assumed $\Delta^+ = 10$,
and attempted to adjust the slope of the DNS data at the wall. 
While the values of the velocity intensities appear reasonable, the shape is evidently incompatible with the DNS data.
On the other hand, figure~\ref{fig:stokes}(b) depicts the predictions of equation~\eqref{eq:bradshaw}, 
again after adjusting the near-wall slope. In this case, the agreement with the DNS data is remarkable, 
particularly for wall distances smaller than the center of the attached eddies (indicated with bullets in the figure). 
This observation serves as plausible evidence that turbulent Stokes layers do indeed exist, 
thereby providing the retardation effects responsible for attenuating the influence of wall-distant eddies.

Although extrapolation is known to be a dangerous, often times ill-posed exercise, we believe that the 
present observations lay down a sufficiently solid background to project today's DNS results to much higher 
Reynolds number than feasible in the near or even foreseeable future. 
When pushed to the extreme, extrapolation of the present data would suggest a scenario where the 
buffer-layer peak of the streamwise velocity variance, while remaining finite, would be much larger than in any current 
DNS and experiments, but roughly at the same inner-scaled position.
Another important feature which we foresee is a secondary distinct peak farther from the wall,
whose strength becomes comparable to and eventually larger than the buffer-layer peak.
The primary caution we emphasize is that these predictions rely on the assumption that the observed 
expansion of the power-law spectrum will persist indefinitely. 
Should a definite $k^{-1}$ spectral range arise at higher Reynolds numbers 
than those achieved thus far in DNS, as suggested by some experimental studies~\citep{baars_20}, 
a logarithmic increasing trend would resume.
Furthermore, it is important to highlight that the current analysis does not offer an explanation 
for the specific value of the spectral scaling exponent in the overlap layer. In contrast, 
\citet{chen_21} rationalized the presence of a $-0.25$ exponent in the proposed defect-power-law 
by attributing it to the flux of additional energy dissipation from the near-wall region to the outer flow.
Thus, further theoretical efforts are still necessary to fully elucidate the observed data and 
unambiguously establish the asymptotic state of wall turbulence. 

\appendix

\section{Box and grid sensitivity analysis}
\label{sec:gridsens}

\begin{figure}
\centerline{
(a) \includegraphics[width=7.0cm]{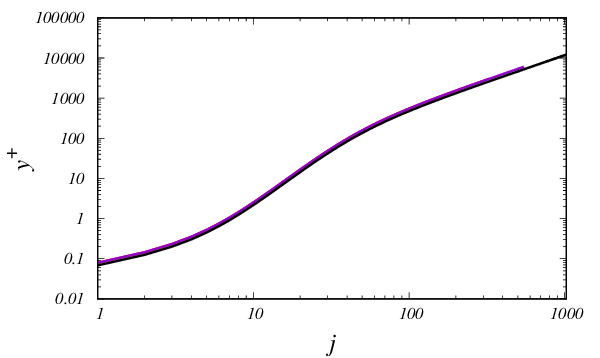}
(b)~\includegraphics[width=7.0cm]{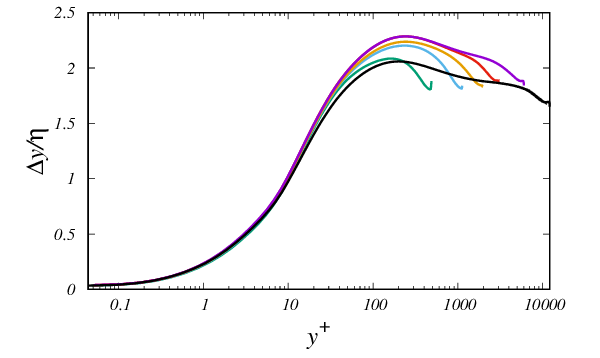}
}
\caption{(a) Distribution of wall-normal grid coordinates as a function of grid index ($j$), and (b) corresponding grid spacings expressed in Kolmogorov units. 
The color codes are as in table~\ref{tab:runs}.}
\label{fig:stretching}
\end{figure}

The grid resolution in the axial and azimuthal directions is decided based on previous experience with 
second-order finite-difference solvers that grid-independent results are obtained provided
$\Delta x^+ \approx 10$, $R^+ \Delta \theta \approx 4.1$~\citep{pirozzoli_21},
hence the associated number of grid points is selected as
$N_z \approx L_z / R \times \Rey_{\tau} / 10$, $N_{\theta} \sim 2 \pi \times \Rey_{\tau} / 4.1$.
The wall-normal distribution of the grid points is designed after the prescriptions set by \citet{pirozzoli_21b},
to resolve the steep near-wall velocity gradients and the local Kolmogorov scale away from the wall, according to
\begin{equation}
y^+(j) = \frac 1{1+(j/j_b)^2} \left[ \Delta y^+_w j + \left( \frac 34 \alpha c_{\eta} j \right)^{4/3} (j/j_b)^2 \right], \label{eq:mapping}
\end{equation}
where $\Delta y^+_w=0.05$ is the wall distance of the first grid point, $j_b=40$ defines the grid index at which transition between the near-wall and the outer mesh stretching takes place, 
and $c_{\eta}=0.8$ guarantees resolution of wavenumbers up to $k_{\max} \eta = 1.5$,
with $\eta$ the local Kolmogorov length scale.
The distributions of the grid points for the DNS listed in table~\ref{tab:runs} are shown in figure~\ref{fig:stretching}(a). A slightly finer mesh is used for flow case G than for the other cases.
The wall-normal resolution is verified a-posteriori in figure~\ref{fig:stretching}(b), where we show 
the grid spacing expressed in local Kolmogorov units, 
which is found to be no larger that $2.2$, throughout the radial direction. 

\begin{figure}
\centering
(a)~\includegraphics[width=6.0cm]{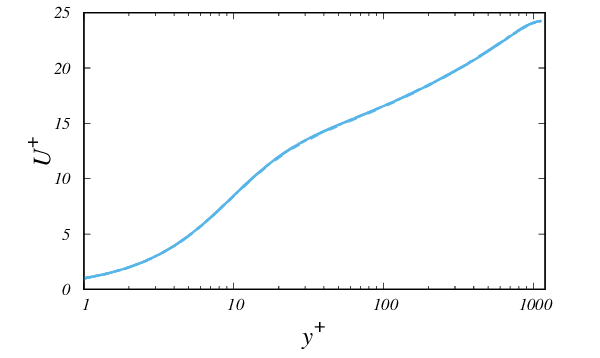} \\
(b)~\includegraphics[width=6.0cm]{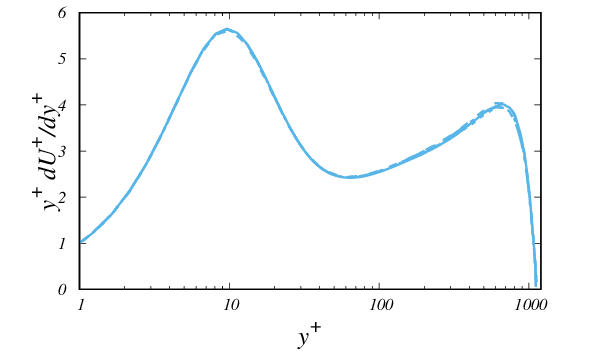} \\
(c)~\includegraphics[width=6.0cm]{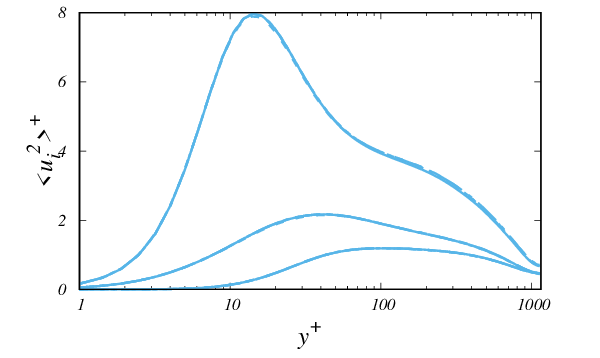}
\caption{Box and grid sensitivity study for one-point statistics: 
(a) inner-scaled mean velocity profiles and streamwise velocity variances, 
(b) log-law diagnostics function,
(c) velocity variances. 
Flow cases C, C-FY, C-L, C-LL are shown, see table~\ref{tab:runs} for the line style.}
\label{fig:uplus_gridsens}
\end{figure}

\begin{figure}
\centering
(a) \includegraphics[width=6.0cm]{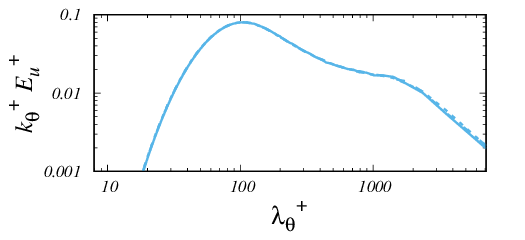} 
(b) \includegraphics[width=6.0cm]{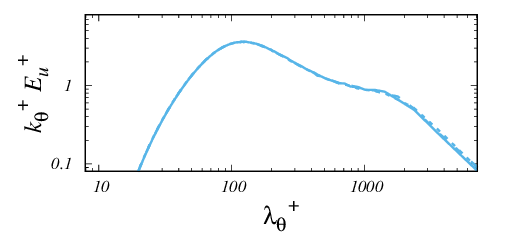} \\
(c) \includegraphics[width=6.0cm]{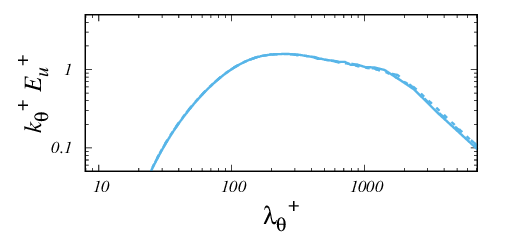} 
(d) \includegraphics[width=6.0cm]{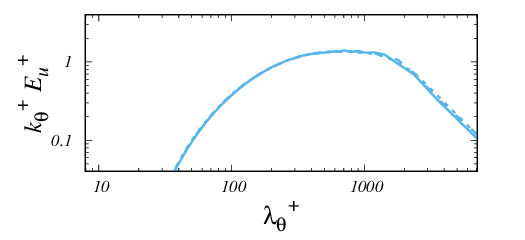} \\
(e) \includegraphics[width=6.0cm]{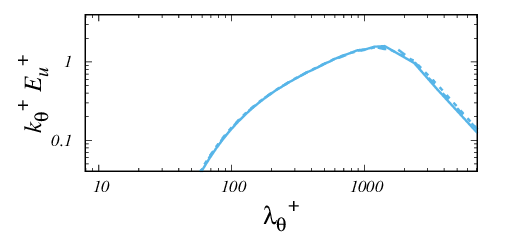}
(f) \includegraphics[width=6.0cm]{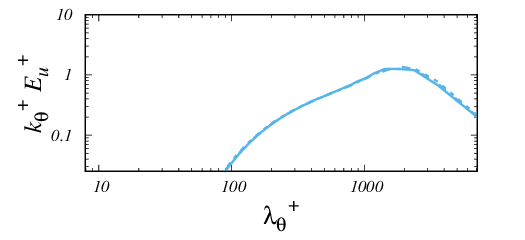} \\
\caption{Box and grid sensitivity study for
pre-multiplied spanwise spectral densities of streamwise velocity at various wall distances:
$y^+=1$ (a), $y^+=15$ (b), $y^+=50$ (c), $y^+=100$ (d), $y^+=200$ (e), $y^+=400$ (f).
Flow cases C, C-FY, C-L, C-LL are shown, see table~\ref{tab:runs} for the line style.}
\label{fig:specz_inner_44000}
\end{figure}

The sensitivity of the computed results to the computational box size and grid resolution 
has been assessed through additional simulations conducted at $\Rey_b = 44000$ (flow case C), 
as listed in table~\ref{tab:runs}. Specifically, we have doubled the number of points in the 
radial direction (flow case C-FY) and doubled and tripled the pipe length (flow cases C-L and C-LL, respectively). 
All these flow cases have been simulated for numerous eddy turnover times, 
for the purpose of removing the time sampling error.
The resulting change in the one-point statistics is well below $1\%$, 
as illustrated in figure~\ref{fig:uplus_gridsens}, even for flow properties 
that are challenging to converge over time, such as the log-law indicator function~\citep{hoyas_23}.
Figure~\ref{fig:specz_inner_44000} additionally presents a comparison of the 
pre-multiplied spanwise spectra at several off-wall positions for the same flow cases. 
Once again, box and grid independence is demonstrated to be excellent, indicating that 
uncertainties primarily stem from finite time sampling.

\section{Mean velocity profiles} 
\label{sec:statistics}

\begin{figure}
 \centerline{
 (a) \includegraphics[width=7.0cm]{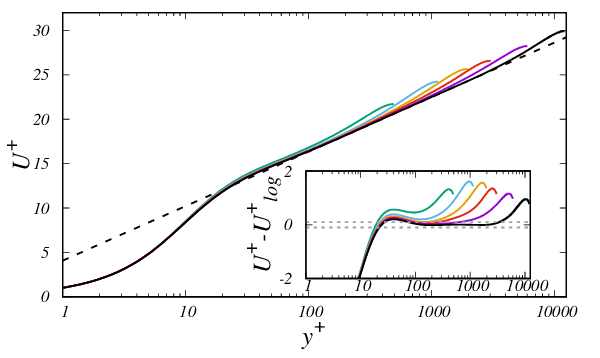}
 (b)~\includegraphics[width=7.0cm]{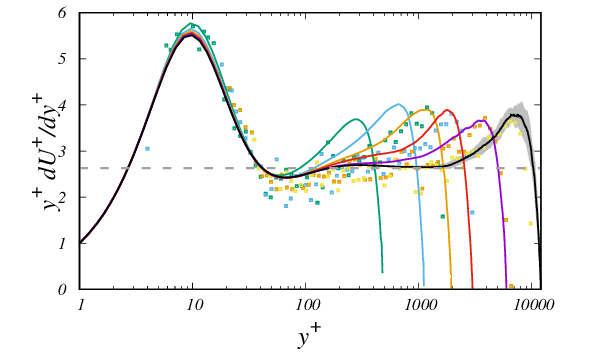}
 }
\caption{Inner-scaled mean velocity profiles obtained from DNS are presented in panel (a), along with 
the corresponding log-law diagnostic function shown in panel (b). In panel (a), 
	the dashed line represents the logarithmic fit $U_{\log}^+ = \log y^+ / 0.38 + 4.3$. 
In panel (b), the dashed horizontal line denotes the inverse of the 
expected K\'arm\'an constant, $\kappa = 0.38$, while symbols denote 
Princeton SuperPipe data~\citep{mckeon_05} at $\Rey_{\tau} = 1825, 3328, 6617, 10914$. 
The shaded grey regions denote the expected range of uncertainty for flow case G 
(not visible in panel (a)). Color codes for the lines are as described in table~\ref{tab:runs}.}
\label{fig:uplus}
\end{figure}

The mean velocity profiles for the DNS cases listed in table~\ref{tab:runs} 
are presented in figure~\ref{fig:uplus}(a), accompanied by the associated log-law diagnostic 
functions shown in panel (b). The latter quantity is commonly used to verify the presence of 
a genuine logarithmic layer in the mean velocity profile, which would be indicated by a plateau.
This diagnostic function exhibits two peaks: one corresponding to the buffer layer, 
which is nearly universal in inner scaling at $\Rey_{\tau} \gtrsim 10^3$, 
and an outer peak corresponding to the wake region, which also becomes approximately 
universal in inner units. Between these two regions, the distribution tends to change 
significantly with the Reynolds number. While previous DNS and analyses suggested 
linear deviations from the 
logarithmic behaviour~\citep{afzal_73,jimenez_07,luchini_17, pirozzoli_21}, 
at the highest Reynolds number achieved in this study, the onset of a genuine logarithmic layer is observed, 
starting at $y^+ \approx 500$ and extending up to $y/R \approx 0.2$. 
This finding is consistent with the SuperPipe data (symbols in panel (b)), 
recent DNS of plane channel flow \citep{hoyas_22}, and in line with the theoretical analysis 
of \citet{monkewitz_21}. The data support a value of the Kármán constant of $\kappa \approx 0.38$, 
slightly less than estimated in previous studies, in which, however, a genuine logarithmic layer was not observed.
The figure reveals a large scatter associated with limited time convergence of flow case D for this indicator, 
confirming recent analyses~\citep{hoyas_23}. While the predictions for the inner layer appear to be robust, 
much longer integration times are required to achieve satisfactory convergence of statistics associated with velocity derivatives. However, achieving such convergence is currently beyond the capabilities of DNS.
Figure~\ref{fig:uplus}(a) confirms that the mean velocity profiles tend to cluster around a common logarithmic distribution. 
The DNS velocity profiles for $\Rey_{\tau} \ge 10^3$ follow this distribution with deviations of no more than $0.1$ 
inner units from $y^+ \approx 30$ to $y/R \approx 0.15$, where the core region develops.

\section{Comparison with DNS of channel flow} \label{sec:channel}

\begin{figure}
\centering
(a)~\includegraphics[width=6.0cm]{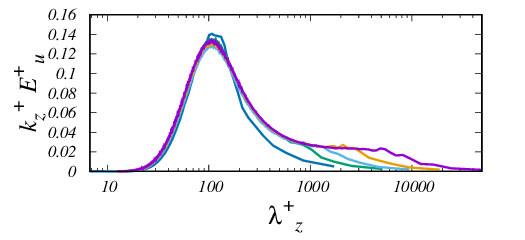}  ~(b)~\includegraphics[width=6.0cm]{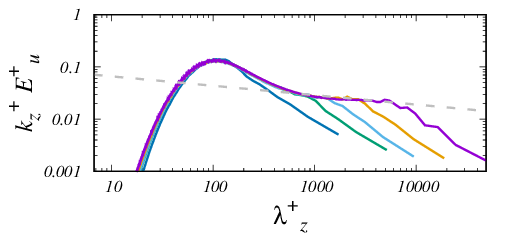} \\
(c)~\includegraphics[width=6.0cm]{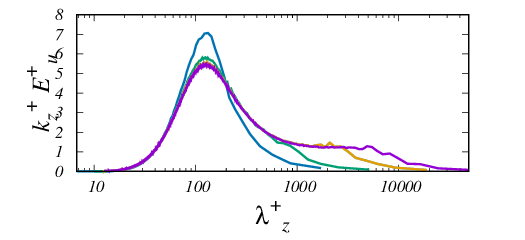} ~(d)~\includegraphics[width=6.0cm]{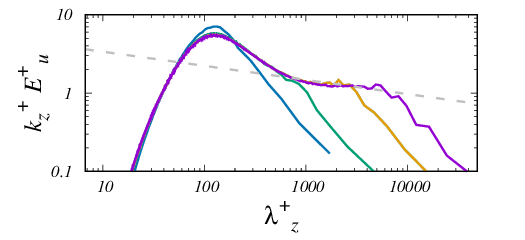} \\
(e)~\includegraphics[width=6.0cm]{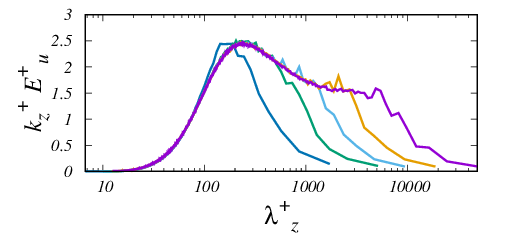} ~(f)~\includegraphics[width=6.0cm]{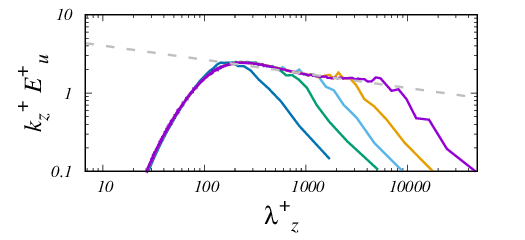} \\
(g)~\includegraphics[width=6.0cm]{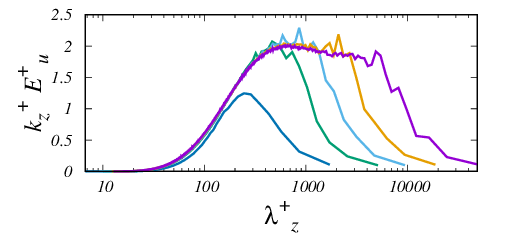}~(h)~\includegraphics[width=6.0cm]{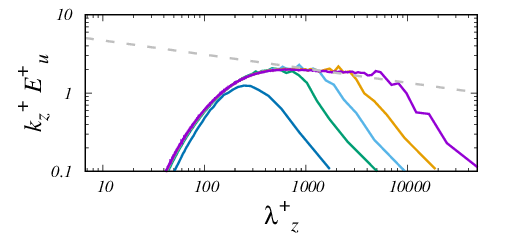} \\
(i)~\includegraphics[width=6.0cm]{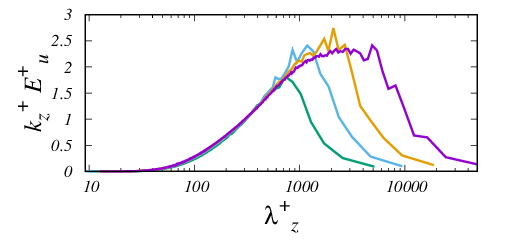}~(j)~\includegraphics[width=6.0cm]{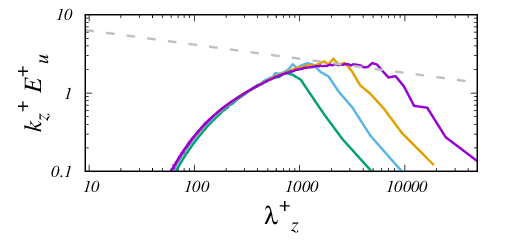} \\
(k)~\includegraphics[width=6.0cm]{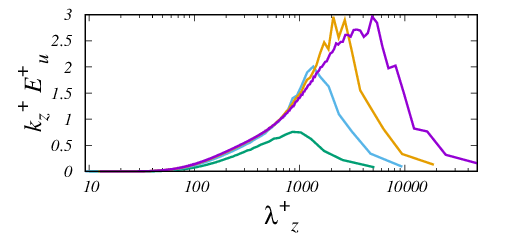}~(l)~\includegraphics[width=6.0cm]{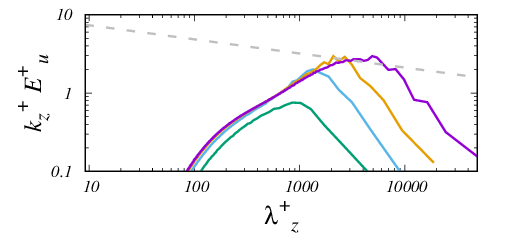} \\
\caption{Pre-multiplied spanwise spectral densities of streamwise velocity from DNS of channel flow~\citep{lee_15}, at various wall distances: $y^+=1$ (a,b), $y^+=15$ (c,d), $y^+=50$ (e,f), $y^+=100$ (g,h), $y^+=200$ (i,j), $y^+=400$ (k,l).  A semi-log representation is used in the left-hand-side panels, and a log-log representation is used in the right-hand-side panels. The color codes correspond to Reynolds numbers $\Rey_{\tau} =$ 180 (blue), 550 (green), 1000 (cyan), 2000 (orange), 5200 (purple).}
\label{fig:specz_inner_LM}
\end{figure}

\begin{figure}
\centerline{
(a) \includegraphics[width=7.0cm]{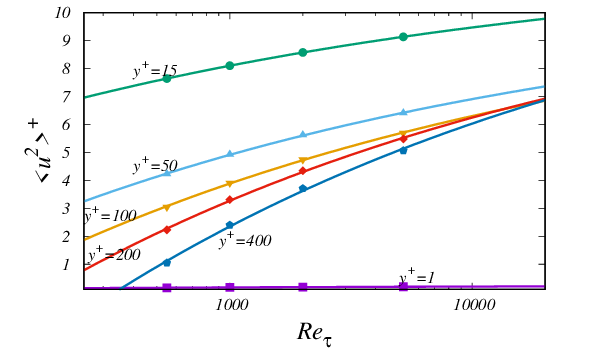}  
(b) \includegraphics[width=7.0cm]{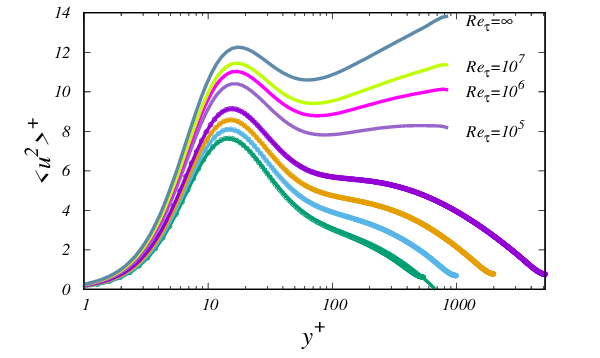}  
}
\caption{
Analysis of velocity variance for turbulent channel flow~\citep{lee_15}:
(a) Streamwise velocity variances (symbols) as a function of $\Rey_{\tau}$ at various off-wall positions, 
along with corresponding fits according to equation \eqref{eq:urms2}; and
(b) Predicted wall-normal distributions of streamwise velocity variances at various $\Rey_{\tau}$, 
again according to equation~\eqref{eq:urms2}. 
In panel (b), the symbols denote the DNS data used to determine the fit coefficients $A(y^+)$ and $B(y^+)$, 
with color codes corresponding to $\Rey_{\tau} =$ 550 (green), 1000 (cyan), 2000 (orange), and 5200 (purple).
}
\label{fig:urms2_LM}
\end{figure}

In figure~\ref{fig:specz_inner_LM}, we present the spanwise spectra 
of the streamwise velocity obtained from the channel flow DNS conducted by~\citet{lee_15}, 
at several off-wall locations, while maintaining a constant $y^+$. 
This figure offers clear evidence that universality of the spectra at small scales 
is achieved when inner scaling is employed, at least for $\Rey_{\tau} \gtrsim 2000$. 
Furthermore, this universal inner-scaled layer tends to become more extended towards 
longer wavelengths as $\Rey_{\tau}$ increases, until a transition to an $h$-scaled spectral range occurs.
These distributions support the findings reported in figure~\ref{fig:specz_inner}, 
indicating that the premultiplied spectra at all considered wall distances exhibit 
a negative-power-law range, with an exponent of approximately $0.18$, as for pipe flow.
However, due to the more limited range of available Reynolds numbers, 
a clear power-law behaviour is visible only up to $y^+ \approx 100$.
Similar to our analysis for pipe flow, figure~\ref{fig:urms2_LM}(a) 
illustrates the Reynolds number trends of the velocity variance at various 
distances from the wall. This figure reaffirms the accuracy of equation \eqref{eq:urms2} 
in fitting the DNS data and visually highlights deviations from a logarithmic behaviour.

Figure~\ref{fig:urms2_LM}(b) depicts the resulting extrapolated distributions 
at extreme Reynolds numbers. The outcome is qualitatively very similar to that 
shown in figure~\ref{fig:urms2_extr} for pipe flow, demonstrating saturation 
of the buffer-layer peak and the onset of a dominant outer-layer peak. 
We estimate that the asymptotic value of the buffer-layer peak is approximately $12.3$, 
while the wall dissipation should asymptote to about $0.27$, 
which aligns well with the numerical values determined for pipe flow.

\backsection[Acknowledgments]{
Discussions with Yongyun Hwang, Peter Monkewitz, Javier Jimenez, Alexander J. Smits, and Tie Wei are gratefully appreciated. I acknowledge that the results reported in this paper have been achieved using the EuroHPC Research Infrastructure resource LEONARDO based at CINECA, Casalecchio di Reno, Italy, under a LEAP grant. Constructive criticisms from the anonymous referees are also gratefully acknowledged.}

\backsection[Funding]{This research received no specific grant from any funding agency, commercial or not-for-profit sectors.}

\backsection[Declaration of interests]{The authors report no conflict of interest.}

\backsection[Data availability statement]{The data that support the findings of this study are openly available at 
the web page http://newton.dma.uniroma1.it/database/}


\bibliographystyle{jfm}
\bibliography{references}
\end{document}